\documentclass[11pt]{article}
\usepackage{cite}
\usepackage{amsmath,amsfonts,amssymb}
\usepackage[small,bf,hang]{caption}
\usepackage{slashed}
\usepackage{mathabx}
\usepackage{latexsym,epsfig}
\usepackage{stmaryrd}

\usepackage[vcentermath]{youngtab}

\usepackage[usenames,dvipsnames]{color}
\usepackage{amssymb}
\usepackage{breqn}

\def\hybrid{
        \topmargin -20pt
        \oddsidemargin 0pt
        \headheight 0pt \headsep 0pt
        \textwidth 6.25in 
        \textheight 9.5in 
        \marginparwidth .875in
        \parskip 5pt plus 1pt \jot = 1.5ex}

\hybrid

 \csname
@addtoreset\endcsname{equation}{section}

\def\moth{\mathsurround=0pt}
\newdimen\zo \zo=0pt

\def\tick{\leaders\hrule height 0.5ex depth 0pt \hskip 0.5pt}
\def\upboxfill{$\moth \setbox\zo\hbox{\tick}%
  \hskip 3pt\hbox to 0pt{$\tick$\hss}\hrulefill \hbox to 7.5pt{$\tick$\hss}$}

\def\dtick{\leaders\hrule height .34pt depth 0.5ex \hskip 0.5pt}
\def\downboxfill{$\moth \setbox\zo\hbox{\dtick}%
  \hskip 2pt\hbox to 0pt{$\dtick$\hss}\hrulefill \hbox to 2pt{$\dtick$\hss}$}

\def\bec{\begin{center}}
\def\ec{\end{center}}

 \def\det{{\rm det\,}}
\def\be{\begin{equation}}
\def\ee{\end{equation}}
\def\bea{\begin{eqnarray}}
\def\eea{\end{eqnarray}}
\def\ba{\begin{array}}
\def\ea{\end{array}}

\allowdisplaybreaks[2]

\begin{document}

\begin{titlepage}
\rightline{}
\rightline{December 2015}
\begin{center}
\vskip 1.6cm
{\Large \bf {Perturbative Double Field Theory 
on General Backgrounds
}}\\
 \vskip 2.0cm
{\large {Olaf Hohm${}^1$ and Diego Marques${}^2$}}
\vskip 0.5cm

{\it {${}^1$Simons Center for Geometry and Physics}}\\
{\it {Stony Brook University}}\\
{\it {Stony Brook, NY 11794-3636, USA}} \\[1ex]
$^2$ {\it Instituto de Astronom\'ia y F\'isica del Espacio (CONICET-UBA)}\\
{\it Ciudad Universitaria, C.P. 1428 }\\ {\it Buenos Aires, Argentina} \\[1ex]

\vskip 0.5cm

{\small \verb"ohohm@scgp.stonybrook.edu, diegomarques@iafe.uba.ar"}

\vskip 1cm
{\bf Abstract}
\end{center}

\vskip 0.2cm

\noindent
\begin{narrower}

{\small

We develop the perturbation theory of double field theory around
arbitrary solutions of its field equations. The exact gauge transformations are
written in a manifestly background covariant way and contain
at most quadratic terms in the field fluctuations.  We expand the generalized
curvature scalar to cubic order in fluctuations and thereby determine the cubic action
in a manifestly background covariant form.
As a first application we specialize this theory to group manifold backgrounds, such as
$SU(2) \simeq S^3$ with $H$-flux.  In the full string theory this corresponds to
a WZW background CFT.
Starting from closed string field theory, the cubic action around such backgrounds has been computed
before by Blumenhagen, Hassler and L\"ust.
We establish precise agreement with the cubic action derived from double field theory.
This result confirms that double field theory is applicable to arbitrary curved background solutions,
disproving assertions in the literature to the contrary. }

\end{narrower}

\vskip 1.5cm

\end{titlepage}

\tableofcontents





\section{Introduction}
Double field theory is the proper framework that makes manifest the global $O(d,d)$
symmetry of the low-energy effective actions of closed string theory implied by T-duality
\cite{Siegel:1993th,Siegel:1993bj,Hull:2009mi,Hull:2009zb,Hohm:2010jy,Hohm:2010pp},
including $\alpha'$ corrections \cite{Hohm:2013jaa,Hohm:2014xsa,Marques:2015vua};
see
\cite{Aldazabal:2013sca,Berman:2013eva,Hohm:2013bwa} for reviews.
The original construction by Hull and Zwiebach started from closed string field theory on a
flat toroidal background with constant metric $G_{ij}$ and Kalb-Ramond field $B_{ij}$ and
determined the cubic action \cite{Hull:2009mi}. By construction, this action depends on the background,
but it can be verified that it is actually background independent in the sense that any constant shift
of the background can be absorbed into a shift of the fluctuation, up to field redefinitions \cite{Hohm:2010jy}.
This property is shared with the full closed string field theory \cite{Sen:1993mh}.
A unique manifestly background independent action can then be constructed, which
is valid to all orders in fields and for arbitrary curved background solutions \cite{Hohm:2010jy}.

In this paper we develop the perturbation theory of double field theory (DFT)
around arbitrary solutions of the field equations. We determine the exact gauge transformations
of the fluctuation fields, and we expand the generalized curvature scalar to cubic order in fields, thereby
determining the cubic action valid for arbitrary background solutions. (The quadratic action was determined
recently in \cite{Ko:2015rha}.)
Employing the geometry of DFT, these results can be written in a manifestly background covariant form,
using covariant derivatives and curvature tensors.
This computation could be done in any formulation of DFT, but it is greatly simplified
by using the frame-like geometry developed by Siegel in \cite{Siegel:1993th}
(and related in \cite{Hohm:2010xe} to the explicit actions of \cite{Hohm:2010jy,Hohm:2010pp}).
In particular, the field variables emerging from the frame field can be identified
with those emerging from closed string field theory around flat space to all orders in
perturbation theory, as was proved in \cite{Hohm:2011dz}.
Here we extend the results of \cite{Hohm:2011dz} by taking the background to be arbitrary.

There are various potential applications for the perturbative formulation of DFT developed
here, ranging from quantum loop computations by the background field method to
cosmological perturbation theory for string cosmology. As a first application
we specialize to backgrounds given by group manifolds with $H$-flux
in order to clarify the cubic theory determined before by Blumenhagen, Hassler and
L\"ust in \cite{Blumenhagen:2014gva}.
Their computation started from the WZW model based on a Lie group G,
which is taken to be the background CFT about which a closed string field theory action
can be evaluated \cite{Zwiebach:1992ie}.
The resulting cubic action takes the same structural form as the original cubic
action by Hull and Zwiebach, however, with all basic structures being deformed
due to the non-flat geometry of the group manifold G.

More specifically, the following new features emerge:
 \textit{i)} the `strong constraint' or `section constraint' originating from the level-matching constraint
 is seemingly deformed, involving derivatives of the background metric;
 \textit{ii)}  the gauge transformations and the gauge algebra are deformed by the structure constants of G; and
 \textit{iii)} the action receives a `potential' term quadratic in the structure constants.
As the basic structures are thus deformed relative to the original cubic theory on which the
background-independent DFT of \cite{Hohm:2010jy} was based,
the results of \cite{Blumenhagen:2014gva} could be (and have been) misinterpreted
as implying that a more general, extended DFT is needed in order to be applicable to certain non-flat backgrounds.
We will show here that no such extension is needed: the cubic action obtained from
the original background independent DFT by
expanding about the appropriate group manifold background
reproduces precisely the cubic action following from
string field theory for WZW
backgrounds.\footnote{We have been informed by the authors of \cite{Blumenhagen:2015zma} that this result
is also implicit in (and compatible with) the computation of their sec.~5 that relates an extended action
to the conventional DFT action.}
We thereby confirm, in particular,  the validity of the cubic couplings determined in \cite{Blumenhagen:2014gva}.

The above conclusion reaffirms the applicability of DFT to arbitrary curved background solutions.
This is as to be expected, because the group manifold backgrounds encoded in WZW models
are fully geometric: they are described by a conventional metric  and $B$-field.
Thus, to leading order in $\alpha'$, the corresponding space-time action is governed by the
usual low-energy action for a metric, $B$-field and dilaton, which in turn is completely captured
by DFT  \cite{Hohm:2010jy}. Moreover, not only is it possible to describe non-flat backgrounds in DFT
(and the closely related `exceptional field theory' \cite{Hohm:2013pua}), it also simplifies significantly their
Kaluza-Klein embedding into the higher-dimensional theory \cite{Hohm:2014qga},
employing the notion of generalized Scherk-Schwarz compactifications \cite{Aldazabal:2011nj,Geissbuhler:2011mx,Lee:2014mla}.
This in fact made it possible recently to resolve long-standing open problems about the consistency of
certain Kaluza-Klein
truncations \cite{Hohm:2014qga,Baguet:2015xha,Baguet:2015sma,Malek:2015hma,Baguet:2015iou}.

In the remainder of
the introduction we summarize some of our key technical
results, in general and applied to WZW backgrounds:

\begin{itemize}

\item[\textit{i)}] We recall that the
 full background independent DFT is subject to the strong constraint
 that for any fields or gauge parameters $X,Y$
  \be\label{STRONG}
  \eta^{MN}\partial_{M}\partial_{N}X \ \equiv \ \partial^{M}\partial_{M}X \ = \ 0\;, \quad
  \partial^MX\,\partial_M Y \ = \ 0\;,
  \qquad
   \eta^{MN} \ = \ \begin{pmatrix}    0 & {\bf 1} \\[0.5ex]
  {\bf 1} & 0 \end{pmatrix}\;,
  \ee
where $\eta_{MN}$ is the $O(D,D)$ invariant metric, and $M,N=1,\ldots, 2D$ are $O(D,D)$ indices.
Upon expanding around a background generalized frame $\bar{E}_{A}{}^{M}$,
where $A=(a,\bar{a})$ is a doubled `Lorentz' index, the `flattened' derivatives
 $D_A \ \equiv \ \bar{E}_{A}{}^{M}  \partial_M$ emerge.
Since the differential operator $\partial^M\partial_M$ is second-order and the background $\bar{E}$ in general
coordinate dependent, the strong constraint generally does \textit{not} take the same form in terms
of the $D_A$.  However, we show that for a large class of backgrounds including
the WZW backgrounds the extra contributions with derivatives of $\bar{E}$ cancel, so that the
constraint reads
 \be\label{flatSTRONG}
  D^{A}D_{A} \ = \ D^a D_a+D^{\bar{a}} D_{\bar{a}} \ = \ 0\;.
 \ee
This agrees with the form of the strong constraint in \cite{Blumenhagen:2014gva}.
More generally, we clarify the nature of the coordinates for WZW backgrounds.
In contrast to the cubic theory on toroidal backgrounds \cite{Hull:2009mi}, there is no physical doubling of coordinates
because there are no winding modes and, accordingly, the proper identification of the physical theory
requires solving the strong constraint (\ref{flatSTRONG}).

\item[\textit{ii)}] The generalized diffeomorphism symmetry of the frame field,
 \be\label{GENLIE}
  \delta_{\xi}E_{A}{}^{M} \ = \ \widehat{\cal L}_{\xi}E_{A}{}^{M} \ \equiv \
  \xi^N\partial_N E_{A}{}^{M}+\big(\partial^M\xi_N - \partial_N\xi^M\big) E_{A}{}^{N}\;,
 \ee
defined in terms of the generalized Lie derivative $\widehat{\cal L}_{\xi}$, gives rise
to gauge transformations of the fluctuation fields, which are given by $h_{a\bar{b}}$
upon employing a particular gauge fixing of the local frame transformations.
The gauge transformations of $h_{a\bar{b}}$ then take the form
  \be\label{INTRObackcovgauge}
  \begin{split}\;
  \delta h_{a\bar{b}} \ = \ \;
   &\bar{\nabla}_{a}\xi_{\bar{b}}-\bar{\nabla}_{\bar{b}}\xi_{a}
   +\xi^{C}\bar{\nabla}_{C}h_{a\bar{b}}+\big(\bar{\nabla}_{a}\xi^{c}-\bar{\nabla}^{c}\xi_{a}\big)h_{c\bar{b}}
   +\big(\bar{\nabla}_{\bar{b}}\xi^{\bar{c}}-\bar{\nabla}^{\bar{c}}\xi_{\bar{b}}\big)h_{a\bar{c}} \\[0.5ex]
   &+h_{a\bar{d}}\big(\bar{\nabla}^{c}\xi^{\bar{d}}-\bar{\nabla}^{\bar{d}}\xi^{c}\big) h_{c\bar{b}}\;,
  \end{split}
 \ee
where $\bar{\nabla}$ are the background covariant derivatives and $\xi_A\equiv \bar{E}_{A}{}^{M}\xi_M$
the flattened gauge parameter. These gauge transformations are
exact, with no higher order terms beyond quadratic order in $h$.
We will show that for the frame field of WZW backgrounds,
which we construct explicitly in terms of the left- and right-invariant Maurer-Cartan forms of G,
the only non-vanishing (generalized) connection components are given by
  \be\label{BackConn}
  \bar{\omega}_{ab}{}^{c} \ = \ \tfrac{1}{3}f_{ab}{}^{c}\;, \qquad
  \bar{\omega}_{\bar{a}\bar{b}}{}^{\bar{c}} \ = \ -\tfrac{1}{3}f_{\bar{a}\bar{b}}{}^{\bar{c}}\;,
 \ee
in terms of the structure constants $f$. Back in (\ref{INTRObackcovgauge})
the gauge transformations then read
 \be
  \delta h_{a\bar{b}} \ = \  \delta^0 h_{a\bar{b}}
  \ + \ \xi^c \,f_{ca}{}^{d} \,h_{d\bar{b}} \ - \  \xi^{\bar{c}}\, f_{\bar{c}\bar{b}}{}^{\bar{d}} \, h_{a\bar{d}}\;,
 \ee
where $\delta^0 h_{a\bar{b}}$ denotes the $f$-independent terms.
These transformations, and the corresponding gauge algebra,
agree precisely with those found in \cite{Blumenhagen:2014gva},
with the deformation in terms of $f$ originating from the background
covariant derivatives.\footnote{A common misconception about DFT is that the appearance of
partial derivatives in the generalized Lie derivative (\ref{GENLIE})
implies that the theory is only consistent on flat space and that
for curved backgrounds the partial derivatives should be replaced by suitably covariantized derivatives.
This is not the case. Covariant derivatives emerge automatically upon expanding
the properly background independent theory
about curved backgrounds.}

\item[\textit{iii)}]
We expand the DFT action written in terms of the dilaton density $e^{-2d}=e^{-2\phi}\sqrt{g}$ and the
generalized curvature scalar ${\cal R}$,
\be
  S_{\rm DFT} \ = \ \int d^{2D}X \,e^{-2d}\,\Big({\cal R}(E,d) \, + \, \lambda\Big)\;,
 \ee
around an arbitrary background.
The cosmological constant $\lambda = -\frac 2 {3 \alpha'} (D-26)$ is non-zero
in order for the group manifolds to be proper string backgrounds.
The resulting action reads schematically
 \be
  S_{\rm DFT} \ = \ \int d^{2D}X\,e^{-2\bar{d}}\Big({\cal L}_0+4\, (1 - 2 d') \bar{\cal R}_{c}{}^{abc}\,h_{a}{}^{\bar{c}}\, h_{b\bar{c}}
  -4\, \bar{\cal R}^{ab\bar{c}d}\, h_{a}{}^{\bar{b}}\, h_{b\bar{c}}\, h_{d\bar{b}} \, + \,  {\cal O}(4) \Big)\;,
 \ee
where $\bar{d}$ is the background dilaton and $d'$ the dilaton fluctuation, and we have only displayed terms involving
explicitly  the background Riemann tensor $\bar{\cal R}$.\footnote{We note that the background covariant
Riemann tensor as well as the covariant derivatives generally require undetermined connection components
which, however, drop out in the full action and gauge transformations.} The remaining
terms denoted by ${\cal L}_0$ are written in terms of the background covariant derivatives $\bar{\nabla}$.
Again, specializing to WZW backgrounds and inserting (\ref{BackConn}), the action
reproduces precisely the cubic action in \cite{Blumenhagen:2014gva}.

\end{itemize}

Summarizing, the original formulation of DFT is sufficient in order to
describe the perturbation theory around arbitrary curved backgrounds in a manifestly
background covariant way. In particular, the cubic theory following from string field theory on WZW backgrounds as determined in
\cite{Blumenhagen:2014gva} is perfectly consistent with the original DFT.
Rather than revealing the limitations of DFT away from non-toroidal backgrounds and showing
where it needs to be extended,
the string field theory computation confirms the applicability of DFT to
curved backgrounds.
One may wonder whether we can learn something from these results about genuinely
non-geometric backgrounds. It is known from investigations of generalized Scherk-Schwarz
compactifications that in general this requires relaxing the strong constraint
\cite{Aldazabal:2011nj,Geissbuhler:2011mx,Dibitetto:2012rk}.\footnote{In a related context, recently the cubic action for closed string theory compactified on a circle at the self-dual radius, with enhanced gauge group $G=SU(2)\times SU(2)$, was computed through scattering amplitudes in \cite{Aldazabal:2015yna} and shown to be
obtainable through a generalized Scherk-Schwarz compactification of DFT, without any need to deform the structure of the parent theory.}
While there are proposals of how to incorporate more general coordinate dependencies
into DFT \cite{Hohm:2011cp,Geissbuhler:2013uka}, we still do not have a proper understanding of
the geometric and physical significance of such truly extended spaces.
Let us also note that  ref.~\cite{Blumenhagen:2014gva} constructs cubic actions
away from the geometric WZW backgrounds, in which a different doubling of coordinates is employed.
We will comment on this in sec.~4.
A more general proposal, put forward for instance in \cite{Cederwall:2014kxa,Blumenhagen:2015zma,Bosque:2015jda},
is to describe the background by a conventional but
doubled geometry, while the physical fields are governed by a generalized geometry.
We will discuss this  and the general issue of background independence
in section~5 and
point out that
such proposals
are problematic in view of the physical requirement of background independence (be it manifest or not).
The rest of this paper is organized as follows. In sec.~2 we briefly review the geometry of DFT
and use this to derive the gauge transformations of the fluctuations.
In sec.~3 we expand the curvature scalar around an arbitrary background and compute
the cubic action. These results are applied in sec.~4 to WZW backgrounds.
We close with some general remarks in sec.~5, while we summarize some explicit formulas for
the cubic couplings in Appendix A and some results for
the simplest WZW background ($S^3$ with $H$-flux) in Appendix B.

\section{DFT symmetries around a background}

\subsection{Generalities of DFT}
We begin by giving a brief review of DFT and the frame-like geometry that will be used in
the following subsections to expand the theory around an arbitrary background solution.
We refer to \cite{Siegel:1993th,Hohm:2010xe} for more details on the frame formulation.
The fundamental fields are given by the dilaton density $e^{-2d}$ and the frame field $E_{A}{}^{M}$,
which are subject to the gauge transformations
  \be\label{DFTgauge}
 \begin{split}
  \delta E_{A}{}^{M} \ &= \ \widehat{\cal L}_{\xi} E_{A}{}^{M}+\Lambda_{A}{}^{B} E_{B}{}^{M} \\[0.2ex]
  \ &\equiv \ \xi^{N}\partial_NE_{A}{}^{M}+\big(\partial^M\xi_N -\partial_N\xi^{M}\big)E_{A}{}^{N}
  +\Lambda_{A}{}^{B} E_{B}{}^{M}\;, \\[0.5ex]
 \delta \big(e^{-2d}\big) \ &= \ \partial_M\big(\xi^Me^{-2d}\big)\;.
 \end{split}
 \ee
Here $\xi^M=(\tilde{\xi}_i,\xi^i)$ is the generalized diffeomorphism parameter w.r.t.~which
we introduced the generalized Lie derivative $\widehat{\cal L}_{\xi}$ whose action on
general $O(D,D)$ tensors is defined analogously.
Moreover, $A=(a,\bar{a})$ is the flat index for the generalized local `Lorentz' group $GL(D)\times GL(D)$, for which
the independent gauge parameters are $\Lambda_{a}{}^{b}$ and $\Lambda_{\bar{a}}{}^{\bar{b}}$.
The frame field is subject to the $GL(D)\times GL(D)$ invariant constraint that
the tangent space metric obtained by `flattening' the $O(D,D)$ metric $\eta_{MN}$ is
block-diagonal,
 \be\label{calGconstr}
  {\cal G}_{AB} \ \equiv \ E_{A}{}^{M} E_{B}{}^{N}\eta_{MN} \ = \ \begin{pmatrix}    {\cal G}_{ab} & 0\\[0.5ex]
  0 & {\cal G}_{\bar{a}\bar{b}} \end{pmatrix}\;.
 \ee
This metric is used to raise and lower flat indices.
The gauge algebra is given by the `C-bracket', i.e., $[\delta_{\xi_1},\delta_{\xi_2}]=\delta_{\xi_{12}}$,
where
 \be\label{Cbracket}
  \xi_{12}^M \ \equiv \ [\, \xi_2, \,\xi_1\, ]^M_{C} \ \equiv \ \xi_2^N\partial_N\xi_1^M- \xi_1^N\partial_N\xi_2^M
  -\tfrac{1}{2}\xi_{2N}\partial^M\xi_1^N   +\tfrac{1}{2}\xi_{1N}\partial^M\xi_2^N  \;.
 \ee
We record for later use that the C-bracket coincides with the antisymmetrization of the
generalized Lie derivative, i.e.,
 \be\label{DorfmanForm}
  [\, V\, , \;W\, ]_{C} \ = \ \tfrac{1}{2}(\widehat{\cal L}_{V}W -\widehat{\cal L}_{W}V\big)\;,
 \ee
for arbitrary generalized vectors $V, W$.
Next, we introduce connections $\omega$ and covariant derivatives $\nabla$
for the $GL(D)\times GL(D)$ gauge symmetry.  Writing $E_{A}=E_{A}{}^{M}\partial_M$ for
the flattened partial derivatives, we define
 \be
  \nabla_AV_B \ = \ E_{A}V_{B}+\omega_{AB}{}^{C} V_{C}\;, \qquad
  \nabla_AV^B \ = \ E_{A}V^B-\omega_{AC}{}^{B}V^C\;.
\ee
One can impose covariant constraints in order to (partially) express the connections in terms of
the frame field and the dilaton \cite{Siegel:1993th,Hohm:2010xe}. There are
undetermined connection components, but they will drop out of the DFT action.
Without repeating the details of this analysis, in the following we simply summarize
the form of the determined connection components and invariant curvatures.
These are most efficiently written in terms of
 \be\label{FDef0}
   {\cal F}_{ABC} \ \equiv \ (\widehat{\cal L}_{{E}_{A}}{E}_{B}{}^{M}){E}_{CM}\;,
  \ee
which are related to the generalized `coefficients of anholonomy' $\Omega_{AB}{}^{C}$ defined by
 \be
  \big[ {E}_{A},{E}_{B}\big]_{C}^M \ \equiv \ \Omega_{AB}{}^{C}{E}_{C}{}^{M}\;,
 \ee
via the partial derivative of the tangent space metric,
 \be
   \Omega_{ABC} \  = \ {\cal F}_{ABC}-\frac{1}{2}E_{C}\,{\cal G}_{AB}\;.
 \ee
Note that by the constraint (\ref{calGconstr}) on ${\cal G}$ we have the special cases
 \be
  \Omega_{a\bar{b}}{}^{\bar{c}} \ = \ {\cal F}_{a\bar{b}}{}^{\bar{c}} \;, \qquad
  \Omega_{\bar{a}b}{}^{c} \ = \ {\cal F}_{\bar{a}b}{}^{c}\;.
 \ee
Using the antisymmetry of $\Omega$ in its first two indices we can also derive
\be\label{FantiREl}
 {\cal F}_{ca\bar{b}} \ = \ -{\cal F}_{c\bar{b}a} \ = \ -\Omega_{c\bar{b}a} \ =  \ \Omega_{\bar{b}ca} \ = \ {\cal F}_{\bar{b}ca}\;,
\ee
and
 \be
 \begin{split}
  {\cal F}_{bac} \ &= \ -{\cal F}_{abc}+E_{c}{\cal G}_{ab}\;, \\[1ex]
  {\cal F}_{bca} \ &= \ {\cal F}_{abc}+E_{b}{\cal G}_{ca}-E_{c}{\cal G}_{ba}\;.
 \end{split}
 \ee
By repeated use of these relations one can then prove
 \be\label{FIDENTITY}
  {\cal F}_{abc} \ = \ {\cal F}_{[abc]}+\frac{1}{2}\big(E_{a}{\cal G}_{bc}-E_{b}{\cal G}_{ca}+E_{c}{\cal G}_{ab}\big)\;.
 \ee
We can finally collect the determined spin connections $\omega_{ABC}$, written in terms of ${\cal F}$
and the tangent space metric. They are given by
 \be\label{connections}
  \begin{split}
   \omega_{a\bar{b}}{}^{\bar{c}} \ &= \ -{\cal F}_{a\bar{b}}{}^{\bar{c}}\;, \qquad\quad
   \omega_{\bar{a}b}{}^{c} \ = \ -{\cal F}_{\bar{a}b}{}^{c}\;. \\[1ex]
   3\, \omega_{[abc]} \ &=  \ -{\cal F}_{[abc]}\;, \quad
   \;\; 3\, \omega_{[\bar{a}\bar{b}\bar{c}]} \ =  \ -{\cal F}_{[\bar{a}\bar{b}\bar{c}]}\;, \\[1ex]
   \omega_{a(bc)} \ &= \ -\tfrac{1}{2}D_{a}{\cal G}_{bc}\;, \quad
   \omega_{\bar{a}(\bar{b}\bar{c})} \ = \ -\tfrac{1}{2}D_{\bar{a}}{\cal G}_{\bar{b}\bar{c}}\;, \\[1ex]
   \omega_{ba}{}^{b} \ &= \ -{\cal F}_{a}\;, \qquad
   \quad \;\; \omega_{\bar{b}\bar{a}}{}^{\bar{b}} \ = \ -{\cal F}_{\bar{a}}\;,
 \end{split}
 \ee
where we introduced the short-hand notation
\be\label{Omegatilde}
 {\cal F}_A \ \equiv \ \partial_M E_A{}^M - 2 E_A d\;.
\ee
The covariant derivatives built with these connections, together with the undermined projections,
transform fully covariantly under local frame
transformations. They also transform covariantly under generalized diffeomorphisms acting on fields with only
flat indices.\footnote{Upon introducing Christoffel-type connections one may also define derivatives that
are covariant acting on arbitrary tensors, see sec.~5.3 of \cite{Hohm:2010xe} and \cite{Hohm:2011si,Jeon:2011cn}
for explicit expressions.}

We close this subsection by discussing the generalized Riemann tensor, from which
a generalized Ricci tensor and a generalized Ricci scalar are constructed
by taking appropriate contractions and projections. The gauge invariant generalized Riemann tensor can be
defined as
\be
{\cal R}_{ABCD} \ = \ \tfrac 1 2 \left(R_{ABCD} + R_{CD AB} - \omega_{EAB} \omega^E{}_{C D}\right)\;,
\ee
where
\be
R_{A B C D} \ = \ E_A \omega_{BCD} -E_B \omega_{ACD} + \omega_{AC}{}^E \omega_{B E D}- \omega_{BC}{}^E \omega_{AED} - \Omega_{A B}{}^E \omega_{ECD}\;.
\ee
From this we define
 \be
  {\cal R} \ \equiv \ 2\,{\cal R}_{ab}{}^{ab} \ = \ -2\,{\cal R}_{\bar{a}\bar{b}}{}^{\bar{a}\bar{b}}\;,
  \qquad
  {\cal R}_{a\bar{b}} \ = \ 2\, {\cal R}_{\bar{c}a\bar{b}}{}^{\bar{c}} \ = \ 2\,{\cal R}_{c\bar{b}a}{}^{c}\;,
 \ee
where we refer to \cite{Siegel:1993th,Hohm:2010xe} for a proof of the equivalence of both definitions for ${\cal R}$.
Again, without repeating the details of this construction,
we give an explicit expression in terms of the above coefficients of anholonomy as
\begin{eqnarray}\label{Curvatures}
{\cal R} &=& -4\big(E_a {\cal F}^a + \tfrac 1 2 {\cal F}_a^2 + \tfrac 1 2 E_a E_b  {\cal G}^{a b} - \tfrac 1 4 \Omega^2_{a b \bar c} - \tfrac 1 {12} {\cal F}_{[abc]}^2 + \tfrac 1 8 E^a {\cal G}^{bc} E_{b} {\cal G}_{a c}\big) \;, \label{GenRicci}\\
{\cal R}_{a \bar b} &=& E_{\bar b} \,{\cal F}_{a} - E_c \,{\cal F}_{\bar b a}{}^c + {\cal F}_{c \bar b}{}^{\bar d} \,
{\cal F}_{\bar d a}{}^c - {\cal F}_{\bar b a}{}^c \, {\cal F}_c\;.
\end{eqnarray}
These expressions are not symmetric in unbarred and barred indices, but using identities implied by
the strong constraint, some of which we will discuss below, it can be brought to a fully democratic form.
The DFT action finally reads
 \be\label{DFTaction0}
  S_{\rm DFT} \ = \ \int d^{2D}X \,e^{-2d}\,\big({\cal R} \, + \, \lambda\big)\;,
 \ee
where $\lambda$  is an arbitrary (cosmological) constant.
Since ${\cal R}$ is a scalar and $e^{-2d}$ a density, the gauge invariance is manifest.
The field equations for the dilaton and the frame field are given by ${\cal R}=-\lambda$ and ${\cal R}_{a\bar{b}}=0$,
respectively.

\subsection{Background expansion}
We now investigate DFT expanded around an arbitrary background solution.
As in \cite{Siegel:1993th,Hohm:2011dz} we expand the frame field so that the fluctuations carry flat indices,
writing
 \be\label{framexapns}
  E_{A}{}^{M} \ = \ \bar{E}_{A}{}^{M}-h_{A}{}^{B}\bar{E}_{B}{}^{M}\;,
 \ee
where $\bar{E}_{A}{}^{M}$ is a general $X$-dependent background solution of the DFT field equations.
We can take this expansion to be exact.
The dilaton is expanded around the background value $\bar{d}$ as
 \be\label{dExp}
  d \  = \ \bar{d} \,  + \, d'\;,
 \ee
where $d'$ is the fluctuation. In the following, when it is clear from the context whether
we consider the full field or the fluctuation, we will drop the prime on $d$.
Moreover, we use the background tangent space metric
$\bar{\cal G}_{AB}\equiv \bar{E}_{A}{}^{M}\bar{E}_{BM}$ to raise and lower flat indices,
and we introduce the notation
 \be
  D_{A} \ \equiv \ \bar{E}_{A}{}^{M}\partial_M\;.
 \ee
In addition, all expressions introduced in the previous subsection for the various
geometric quantities have a background counterpart in this and the following sections,
generically indicated by a bar or straight letters. For instance,  for the ${\cal F}_{ABC}$ defined in (\ref{FDef0})
we write for the background version
  \be\label{FDef}
   F_{ABC} \ \equiv \ (\widehat{\cal L}_{\bar{E}_{A}}\bar{E}_{B}{}^{M})\bar{E}_{CM}\;,
  \ee
which can be computed to be
 \be\label{FForm}
  F_{ABC} \ = \ 3\,D_{[A}\bar E_{B}{}^{M}\,\bar E_{C]M}
  +\frac{1}{2}\big(D_{A}\bar{\cal G}_{CB}+D_{C}\bar{\cal G}_{AB}-D_{B}\bar{\cal G}_{AC}\big)\;.
 \ee
It is straightforward to prove from the definition (\ref{FDef}) that the commutator
of flattened derivatives is given by
 \be\label{DCommu}
  \big[D_A,D_B\big] \ = \ F_{AB}{}^{C}D_{C}\;.
 \ee
We also define
  \be\label{BackFA}
  { F}_A \ \equiv \ \partial_M \bar{E}_A{}^M - 2 D_A \bar{d}\;.
 \ee
 The explicit expressions (\ref{connections}) for the connections then apply equally
to the background connections, just with ${\cal F}$  replaced by $F$, etc.
We denote the corresponding background covariant derivatives by $\bar{\nabla}$
and the background curvatures by $\bar{\cal R}$.

We close this subsection by stating and discussing further identities
satisfied by the background structures. Specifically, we have the
Bianchi-type identities\footnote{Here, these identities are given for the background objects,
but completely analogous relations hold for the full quantities in the background independent theory.}
\begin{eqnarray}\label{FIdentities}
D_A D^A + F_A \,D^A &=& 0\;, \\ \label{SecondIdentities}
D_{[A}F_{BCD]} - \frac 3 4 F_{[AB}{}^E F_{CD]E} &=& 0\;,  \\
2 D_{[A} F_{B]} - D_C F_{A B}{}^C - F_C F_{AB}{}^C &=& 0\;. \label{LastFIdentities}
\end{eqnarray}
These can be verified by straightforward computations using the strong constraint.
As an illustration let us discuss the first relation, which we will discuss further for WZW models.
 To this end, we
have to translate the standard constraints,
 \be
  \partial^M\partial_M X \ = \ 0\;, \qquad
  \partial^MX\,\partial_MY \ = \ 0\;,
 \ee
which hold for arbitrary $X, Y$,
into flat indices. For the second form we immediately get the flattened
 version
  \be
   D^A X\,D_AY \ = \ 0\;,
  \ee
by expressing the derivatives in terms of flattened background derivatives.
For the first form, however, one obtains extra contributions,
 \be\label{flattenedStrong}
 \begin{split}
  \partial^M\partial_M X \ &= \ \bar{E}^{A M}D_{A}\big(\bar{E}^{B}{}_{M} D_{B}X\big) \ = \
  \bar{E}^{A M}\bar{E}^{B}{}_{M} D_{A}D_{B}X+\bar{E}^{A M}D_{A}\bar{E}^{B}{}_{M} D_{B}X\\[1ex]
  \ &= \ D^{A}D_{A}X + D_A \bar {\cal G}^{A B} D_B X -D_{A}\bar{E}^{A M}\bar{E}^{B}{}_{M} D_{B}X \;.
 \end{split}
 \ee
In the second line, the second term can be used to change the differential operator into $D_AD^A$.
The last term is non-vanishing in general. However, rewriting this term with (\ref{BackFA}) in the form $-D_{A}\bar{E}^A{}_{M} =F_M+2\partial_M\bar{d}$,
the dilaton dependent term drops out by the strong constraint, and we obtain
 \be
   \partial^M\partial_M X \ = \   D_{A}D^{A}X +F_M\, \bar{E}^{B M}D_B X\;.
 \ee
This proves eq.~(\ref{FIdentities}).
For the special case $F_A=0$ and constant $\bar {\cal G}$, which we will see to be satisfied for WZW backgrounds,
we conclude
 \be\label{flattConstr}
  D^{A}D_{A} \ = \ D_A D^A \ = \ 0\;,
 \ee
which coincides with the constraint used in \cite{Blumenhagen:2014gva}.

 Finally we turn to another, quadratic identity for the $F_{ABC}$, that
 holds whenever the tangent space metric is constant and $F_A=0$.
 We compute from (\ref{FForm})
 \be
 \begin{split}
  F^{ABC} F_{ABC} \ &= \ 3\big(D^A \bar{E}^{B}{}_{M}\,\bar{E}^{CM}+D^C \bar{E}^{A}{}_{M}\,\bar{E}^{BM}
  +D^{B} \bar{E}^{C}{}_{M}\,\bar{E}^{AM}\big)
  D_{A} \bar{E}_{B}{}^{N} \bar{E}_{CN} \\
  \ &= \ -6\, \partial_N \bar{E}^{BM}\,\partial_M \bar{E}_{B}{}^{N}\;,
 \end{split}
 \ee
where we used the strong constraint.  This is generally non-zero, but it does vanish
for the special case $F_A=0$, which implies $\partial_M \bar{E}_{A}{}^{M}=2D_{A}\bar{d}$.
To prove this, we use the latter condition and the constancy of the $O(D,D)$ metric,
 \be
 \begin{split}
  0 \ = \ \partial_M\partial_N(\bar{E}^{BM} \bar{E}_{B}{}^{N}) \ &= \
  2\,\partial_M\partial_N \bar{E}^{BM}\,\bar{E}_{B}{}^{N}+\partial_M\bar{E}^{BM}\partial_N \bar{E}_{B}{}^{N}
  +\partial_N\bar{E}^{BM}\,\partial_M\bar{E}_{B}{}^{N} \\
  \ &= \ 4D_{B}D^{B}\bar{d}+4D^{B}\bar{d} \, D_B \bar{d} +\partial_N\bar{E}^{BM}\partial_M \bar{E}_{B}{}^{N} \\
  \ &= \ \partial_N\bar{E}^{BM}\partial_M \bar{E}_{B}{}^{N} \;,
 \end{split}
 \ee
using that the strong constraint implies $D^A D_A=0$ under the same assumptions.
We thus conclude
 \be\label{Fsquareconst}
   F^{ABC} F_{ABC} \ \equiv \ F^{abc} F_{abc}+F^{\bar{a}\bar{b}\bar{c}}F_{\bar{a}\bar{b}\bar{c}} \ = \ 0\;,
 \ee
so  that we can write more symmetrically
 \be
  F^{abc} F_{abc} \ = \ \frac{1}{2}(   F^{abc} F_{abc}-F^{\bar{a}\bar{b}\bar{c}}F_{\bar{a}\bar{b}\bar{c}})\;.
 \ee
Note that the unusual sign is due to the indefinite signature of the tangent space
metric (\ref{calGconstr}).

\subsection{Gauge symmetries around arbitrary backgrounds}

Our goal in the rest of this section is to determine the symmetry variations acting on
the fluctuations $h_{AB}$ and $d'$ and to write them in background covariant form.
Let us first note that taking the gauge parameters $\bar{\xi}^M$ and $\bar{\Lambda}_A{}^B$ to be of zero order
in the background expansion
(as indicated by bars), the gauge variations of (\ref{framexapns}) and  (\ref{dExp}) yield
for the background
 \be\label{backgroundVAR}
 \begin{split}
  \delta \bar{E}_{A}{}^{M} \ &= \ \widehat{\cal L}_{\bar{\xi}}\bar{E}_{A}{}^{M}
  +\bar{\Lambda}_{A}{}^{B}\bar{E}_{B}{}^{M}\;, \\[0.5ex]
  \delta \bar{d} \ &= \ \bar{\xi}^N\partial_N\bar{d}-\tfrac{1}{2}\partial_N\bar{\xi}^N\;.
 \end{split}
 \ee
In general this should not be interpreted as an additional physical symmetry because
the background is not allowed to transform. If, however, there exist $\bar{\xi}^M$ and $\bar{\Lambda}_{A}{}^{B}$
so that the above right-hand sides are zero, these are global symmetries (generalized Killing symmetries),
acting non-trivially on the physical (fluctuation) fields. Let us compute their symmetry variations,
starting with the dilaton. Shifting $\xi^M\rightarrow \bar{\xi}^M+\xi^M$, where the new $\xi^M$
represents the first-order part, we obtain from (\ref{DFTgauge}),
 \be
  \delta_{\bar{\xi}}\bar{d} + \delta_{\bar{\xi}}d' + \delta_{\xi}d' \ = \  \big(\bar{\xi}^N+\xi^N\big)\partial_N\big(\bar{d}+d'\big)
  -\tfrac{1}{2}\partial_N\big(\bar{\xi}^N+\xi^N\big)\;,
 \ee
taking the background dilaton $\bar{d}$ to be invariant under the first-order variations w.r.t.~$\xi^M$.
Expanding both sides and using the second equation in (\ref{backgroundVAR}) we read off
 \be
 \begin{split}
  \delta_{\bar{\xi}}d' \ &= \ \bar{\xi}^N \partial_N d'\;, \\[0.5ex]
  \delta_{\xi}d' \ &= \ \xi^N\partial_N\bar{d}+\xi^N\partial_Nd' -\tfrac{1}{2}\partial_N\xi^N\;.
 \end{split}
 \ee
We infer from the first equation that under background diffeomorphisms $d'$ actually
transforms as a scalar.
Next, let us rewrite the first-order gauge transformations of $d'$ in
background covariant form. Flattening the indices with the background frame field in the second
equation above, we obtain
 \be\label{dilatongauge}
  \begin{split}
   \delta_{\xi}d' \ &= \
   \xi^AD_A\bar{d}+ \xi^AD_Ad'-\tfrac{1}{2}\partial_N\big(\xi^A\bar{E}_{A}{}^{N}\big) \\[0.5ex]
   \ &= \  \xi^AD_Ad' -\tfrac{1}{2}\big(D_{A}\xi^A+{F}_{A}\xi^A\big)  \\[0.5ex]
   \ &= \  \xi^A\bar{\nabla}_Ad'-\tfrac{1}{2}\bar{\nabla}_A\xi^A  \;,
  \end{split}
 \ee
where we identified the trace parts of
the background connections, c.f.~the last line in (\ref{connections}) and (\ref{Omegatilde}),
and we used that since $d'$ is a scalar under background diffeomorphisms, $\bar{\nabla}_Ad'=D_A d'$.

Next, we turn to the symmetry transformations of the fluctuations $h_{AB}$.
First, one finds in complete analogy of the above discussion that under background generalized
diffeomorphisms and background frame transformations w.r.t.~$\bar{\Lambda}_A{}^B$,
 \be\label{BACkgroundFRame}
  \bar{\delta} h_{AB} \ = \ \bar{\xi}^N\partial_N h_{AB}+\bar{\Lambda}_A{}^{C} h_{CB}
  +\bar{\Lambda}_B{}^{C} h_{AC}\;.
 \ee
Thus, $h_{AB}$ transforms as a scalar under background
diffeomorphisms and as a 2-tensor under background frame transformations, as indicated
by its index structure.
In the following we focus on the non-trivial part of the gauge symmetries,
setting the background parameters to zero and treating
the gauge parameters $\xi^{M}$ and $\Lambda_{A}{}^{B}$ to be of first order.
Again, in this computation the background is taken
to be invariant,
$\delta\bar{E}_{A}{}^{M}=0$.\footnote{Note, however, that in general
$\widehat{\cal L}_{\xi}\bar{E}_{A}{}^{M}\neq 0$.}
Inserting the expansion (\ref{framexapns}) into the gauge transformations (\ref{DFTgauge})
we then obtain
 \be
 \begin{split}
  -\delta h_{A}{}^{B}\bar{E}_{B}{}^{M} \ = \  &\; \xi^{N}\partial_N( \bar{E}_{A}{}^{M}-h_{A}{}^{B}\bar{E}_{B}{}^{M})
  +\big(\partial^M\xi_N -\partial_N\xi^{M}\big)( \bar{E}_{A}{}^{N}-h_{A}{}^{B}\bar{E}_{B}{}^{N})\\[0.5ex]
  &+\Lambda_{A}{}^{B} ( \bar{E}_{B}{}^{M}-h_{B}{}^{C}\bar{E}_{C}{}^{M})\;.
 \end{split}
 \ee
Multiplying by the inverse background vielbein this reads
 \be\label{deltahAB}
  \delta h_{AB} \ = \ -(\widehat{\cal L}_{\xi}\bar{E}_{A}{}^{M})\bar{E}_{BM}+
  \xi^{N}\partial_N h_{AB}-h_{A}{}^{C}(\widehat{\cal L}_{\xi}\bar{E}_{BM})\bar{E}_{C}{}^{M}
  -\Lambda_{AB}+\Lambda_{A}{}^{C} h_{CB}\;.
 \ee
To simplify this we note with a quick computation, using the generalized Lie derivative (\ref{DFTgauge}),
 \be
  (\widehat{\cal L}_{\xi}\bar{E}_{A}{}^{M})\bar{E}_{BM} \ = \ (\widehat{\cal L}_{\xi^C\bar{E}_{C}}
  \bar{E}_{A}{}^{M})\bar{E}_{BM}
  \ = \ \xi^{C}(\widehat{\cal L}_{\bar{E}_{C}}\bar{E}_{A}{}^{M})\bar{E}_{BM}-D_{A}\xi^{C}\bar{\cal G}_{CB}
  +D_{B}\xi^{C}\bar{\cal G}_{CA}\;.
 \ee
Thus, in terms of (\ref{FDef}) we have
 \be
   (\widehat{\cal L}_{\xi}\bar{E}_{A}{}^{M})\bar{E}_{BM} \ = \
   -D_{A}\xi^{C}\bar{\cal G}_{CB} +D_{B}\xi^{C}\bar{\cal G}_{CA}+\xi^{C} F_{CAB}\;.
 \ee
 Organizing the terms in (\ref{deltahAB}) we obtain the gauge transformations
  \be\label{semifinaldeltah}
   \begin{split}
    \delta h_{AB} \ = \ &\; D_{A}\xi^{C}\bar{\cal G}_{CB}-D_{B}\xi^{C}\bar{\cal G}_{CA}-\Lambda_{AB}
    +\Lambda_{A}{}^{C} h_{CB} \\[1ex]
    &+\xi^{C}D_{C}h_{AB}+D_{B}\xi^{C} h_{AC}-D^{C}\xi^{D}\bar{\cal G}_{DB} h_{AC}\\[1ex]
    &-\xi^{C} F_{CAB}-\xi^{D} F_{DB}{}^{C} h_{AC}\;.
   \end{split}
  \ee
Note that in general the tangent space metric $\bar{\cal G}$ is $X$-dependent and so
cannot be moved inside partial derivatives.

We next discuss an important simplification of the perturbation theory that employs
a convenient gauge fixing of the $GL(D)\times GL(D)$ symmetry. In order to motivate
this gauge condition, note that to lowest order the frame transformations
act as a St\"uckelberg symmetry on the `diagonal' components,
 \be
   \delta h_{ab} \ = \ -\Lambda_{ab}\;, \qquad \delta h_{\bar{a}\bar{b}} \ =  \ -\Lambda_{\bar{a}\bar{b}}\;, \qquad
   \delta h_{a\bar{b}} \ = \ \delta h_{\bar{a}b} \ = \ 0\;.
 \ee
Thus, we can choose a gauge for which
 \be
  h_{ab} \ = \ h_{\bar{a}\bar{b}} \ = \ 0\;. \label{GaugeFixing}
 \ee
This condition has consequences from the constraint (\ref{calGconstr}) stating that the tangent space
metric is block-diagonal,
 \be
  {\cal G}_{a\bar{b}} \ = \ (\bar{E}_{a}{}^{M}-h_{a}{}^{\bar{c}}\bar{E}_{\bar{c}}{}^{M})
  (\bar{E}_{\bar{b}}{}^{N}-h_{\bar{b}}{}^{d}\bar{E}_{d}{}^{N})\eta_{MN} \ = \
  -h_{\bar{b}a}-h_{a\bar{b}} \ = \ 0\;.
 \ee
From this we conclude that $h_{a\bar{b}}=-h_{\bar{b}a}$ describes the independent physical fields
encoded in the frame field. This is as it should be, for it describes $D^2$ degrees of freedom,
the sum of the metric and $b$-field fluctuations, which are the only physical fields
besides the dilaton.

The above gauge fixing requires compensating local  $GL(D)\times GL(D)$ transformations, because
a general diffeomorphism with parameter $\xi^M$ does not preserve the gauge condition.
One quickly finds with (\ref{semifinaldeltah})
that the needed gauge parameter takes the form
 \be
  \Lambda_{ab} \ = \ D_{a}\xi^{c}\,\bar{\cal G}_{cb}-D_{b}\xi^{c}\,\bar{\cal G}_{ca}
  +D_{b}\xi^{\bar{c}}\,h_{a\bar{c}}-D^{\bar{c}}\xi^{d}\,\bar{\cal G}_{db}\,h_{a\bar{c}}
  -\xi^{C} F_{Cab}-\xi^{D} F_{Db}{}^{\bar{c}}\,h_{a\bar{c}}\;,
 \ee
and similarly for the barred one.
Inserting this gauge parameter into (\ref{semifinaldeltah}), specialized to external indices $a\bar{b}$,  one finds
 \be\label{DELTAhSTEP}
  \begin{split}
   \delta h_{a\bar{b}} \ = \ \;&D_{a}\xi^{\bar{c}}\,\bar{\cal G}_{\bar{c}\bar{b}}- D_{\bar{b}}\xi^{c}\,\bar{\cal G}_{ca}
   -\xi^{C}F_{Ca\bar{b}} \\[1ex]
   &+\xi^{C} D_{C} h_{a\bar{b}}+D_{\bar{b}}\xi^{\bar{c}}\,h_{a\bar{c}}-D^{\bar{c}}\xi^{\bar{d}}\,\bar{\cal G}_{\bar{d}\bar{b}}\,
   h_{a\bar{c}}+D_a\xi^{c}\,h_{c\bar{b}}-D^{c}\xi^{d}\,\bar{\cal G}_{da}\,h_{c\bar{b}}\\[0.5ex]
   &-\xi^{D} F_{D\bar{b}}{}^{\bar{c}}\,h_{a\bar{c}}-\xi^{D} F_{Da}{}^{c}\,h_{c\bar{b}}\\[1ex]
   &+D^c\xi^{\bar{c}}\,h_{a\bar{c}} h_{c\bar{b}}-D^{\bar{c}}\xi^{c}\,h_{a\bar{c}}h_{c\bar{b}}
   -\xi^{D} F_{D}{}^{c\bar{c}}\,h_{a\bar{c}} h_{c\bar{b}}\;,
  \end{split} 
 \ee
where the terms are organized in increasing powers of $h$. This is the final form of the
gauge transformations for $h_{a\bar{b}}$. Note, in particular, that we obtain at most terms
quadratic in $h$ to all orders in perturbation theory.
This was established for the expansion around flat space in \cite{Hull:2009zb}, but now we see that this is true
also for the expansion around a general background.

The above gauge transformations are the complete transformations for arbitrary backgrounds,
and it is illuminating to rewrite them in a form that is manifestly covariant under generalized diffeomorphism
transformations of the background. We will show in the following that the above gauge transformations
can be written in terms of the background  covariant derivatives.
To illustrate this, let us focus first on the
inhomogeneous, field-independent terms in the first line of (\ref{DELTAhSTEP}).
We find, using the antisymmetry relations (\ref{FantiREl}) and identifying the connection
components (\ref{connections}),
 \be
 \begin{split}
  D_{a}\xi^{\bar{c}}\,\bar{\cal G}_{\bar{c}\bar{b}}- D_{\bar{b}}\xi^{c}\,\bar{\cal G}_{ca}
   -\xi^{c}F_{ca\bar{b}} -\xi^{\bar{c}}F_{\bar{c}a\bar{b}}
  \ &= \  (D_{a}\xi^{\bar{c}}+F_{a\bar{d}}{}^{\bar{c}}\xi^{\bar{d}})\bar{\cal G}_{\bar{c}\bar{b}}
  -(D_{\bar{b}}\xi^{c} +F_{\bar{b}d}{}^{c}\xi^{d})\bar{\cal G}_{ca}\\[1ex]
  \ &= \ \bar{\nabla}_a\xi^{\bar{c}}\,\bar{\cal G}_{\bar{c}\bar{b}} -\bar{\nabla}_{\bar{b}}\xi^{c}\,\bar{\cal G}_{ca}\\[1ex]
   \ &= \
  \bar{\nabla}_a\xi_{\bar{b}}-\bar{\nabla}_{\bar{b}}\xi_{a}\;.
 \end{split}
 \ee
Here we used in the last step that the background tangent space metric is covariantly constant and
can hence be moved inside the covariant derivative to lower the index on the gauge parameter.

Similarly, for the terms linear and quadratic in $h$ one may replace the flattened partial
derivatives by covariant derivatives, adding and subtracting the necessary connection terms. Using identities
such as (\ref{FIDENTITY}) one can verify that the connection terms precisely cancel against
the explicit $F$ terms in (\ref{DELTAhSTEP}).
Importantly, undetermined connections are needed, but one may verify
that they drop out in the full expression.
The complete final gauge transformations then take the form
 \be\label{backcovgauge}
  \begin{split}\;
  \delta h_{a\bar{b}} \ = \ \;
   &\bar{\nabla}_{a}\xi_{\bar{b}}-\bar{\nabla}_{\bar{b}}\xi_{a}\\[1ex]
   &+\xi^{C}\bar{\nabla}_{C}h_{a\bar{b}}+\big(\bar{\nabla}_{a}\xi^{c}-\bar{\nabla}^{c}\xi_{a}\big)h_{c\bar{b}}
   +\big(\bar{\nabla}_{\bar{b}}\xi^{\bar{c}}-\bar{\nabla}^{\bar{c}}\xi_{\bar{b}}\big)h_{a\bar{c}} \\[1ex]
   &+h_{a\bar{d}}\big(\bar{\nabla}^{c}\xi^{\bar{d}}-\bar{\nabla}^{\bar{d}}\xi^{c}\big) h_{c\bar{b}}\;.
  \end{split}
 \ee
Comparing with the complete perturbative DFT gauge transformations around flat space given in
\cite{Hull:2009zb}, we infer that they take precisely the same form, except that the
partial derivatives are replaced by covariant derivatives.

Let us finally discuss the gauge algebra of these perturbative gauge transformations.
The gauge algebra is changed compared to the background independent (\ref{Cbracket})
because the gauge parameters are redefined. This redefinition is simply given by the
flattening of the vector index, which implies that the gauge algebra is redefined in the same way,
 \be
  \xi_{12}^A \ \equiv \ \xi_{12}^M \, \bar{E}_{M}{}^{A} \ = \ \tfrac{1}{2}\big(\widehat{\cal L}_{\xi_2}\xi_1^M
  -\widehat{\cal L}_{\xi_1}\xi_2^M\big)\bar{E}_{M}{}^{A}\;,
 \ee
where we used the form (\ref{DorfmanForm}) for the C-bracket.
Expanding the gauge parameters in the frame basis $\bar{E}_{A}{}^{M}$ this becomes
 \be
 \begin{split}
   \xi_{12}^A \ &= \ \tfrac{1}{2}\big(\xi_2^B\bar{E}_{B}{}^{N}\partial_N(\xi_1^C\bar{E}_{C}{}^{M})
   +(\partial^M(\xi_2^B\bar{E}_{BN})-\partial_N(\xi_2^B\bar{E}_{B}{}^{M}))\xi_1^C\bar{E}_{C}{}^{N}
   -(1\leftrightarrow 2)\big)\bar{E}_{M}{}^{A} \\[0.5ex]
   \ &= \ \tfrac{1}{2}\big(\xi_2^B\xi_1^C F_{BC}{}^{A}+2\,\xi_2^BD_{B}\xi_1^A-\xi_{2B}D^A\xi_1^B-(1\leftrightarrow 2)\big)\;,
 \end{split}
 \ee
where we used the definition (\ref{FDef}).
Rearranging terms, we have thus shown
\begin{eqnarray}
\xi_{12}^A &=& \xi_2^B D_B \xi_1^A -\xi_1^B D_B \xi_2^A
- \frac 1 2 \xi_{2B} D^A \xi_{1}^{B} + \frac 1 2 \xi_{1B} D^A \xi_{2}^{B}
 + F_{[BC]}{}^{A} \xi_2^B \xi_1^C  \;. \label{deformedbracket}
\end{eqnarray}
We infer that the algebra of gauge symmetries expanded around a background
carries a `deformation' characterized by the background `structure constants'
$F_{AB}{}^{C}$. This will become important in sec.~4 when we analyze WZW backgrounds.

\section{Background field expansion of the action}

\subsection{Expansion of generalized Ricci scalar}
In this section we will expand the background independent DFT (\ref{DFTaction0})
about an arbitrary background and find the effective theory for the perturbations to cubic order.
The generalized Ricci scalar in the frame-like formalism of DFT was given in (\ref{GenRicci}),
\be
{\cal R} = -4\, \left(E_a {\cal F}^a + \frac 1 2 {\cal F}_a^2 + \frac 1 2 E_a E_b {\cal G}^{a b} - \frac 1 4 \Omega_{ab \bar c}^2 - \frac 1 {12} \Omega_{[abc]}^2 + \frac 1 8 E^a {\cal G}^{b c} E_b {\cal G}_{a c} \right)\ . \label{FullGenRicci}
\ee
Using the expansion ansatz introduced in the previous section, we split the generalized frame and dilaton as
\be
E_A{}^M \ = \ \Phi_A{}^B \bar E_B{}^M \ , \ \ \ \ d \ = \ \bar d + d'\ ,
\ee
in background pieces $\bar E_{A}{}^M$ and $\bar d$ and fluctuations around them, $\Phi_A{}^B = \delta_A{}^B - h_A{}^B$  and $d'$. This in turn allows us to decompose the following quantities that enter the action
\begin{eqnarray}
{\cal F}_A &=& \widehat {\cal F}_A + \Phi_A{}^B F_B \nonumber \\
\Omega_{A B C} &=&  \widehat \Omega_{A B C} + \Phi_A{}^D\Phi_B{}^E\Phi_C{}^F \left(F_{DEF} - \frac 1 2 D_F \bar {\cal G}_{D E}\right)  \ , \label{splitting}
\end{eqnarray}
where the hats indicate that the objects depend only on derivatives of the fluctuations,
\begin{eqnarray}
\widehat {\cal F}_A &=& D_B \Phi_A{}^B - 2 \Phi_A{}^B D_B d' \\
\widehat \Omega_{A B C} &=& 2 \Phi_{[A|}{}^D D_D \Phi_{|B]}{}^E \bar {\cal G}_{E F} \Phi_C{}^F  + \Phi_C{}^D D_D \Phi_{[A}{}^E \Phi_{B]}{}^F \bar {\cal G}_{E F} \ .
\end{eqnarray}
In this section we will everywhere assume that the background metric $\bar {\cal G}_{A B}$ is constant,
which can be viewed as a gauge fixing condition for the background $GL(D)\times GL(D)$ frame
transformations.  This in turn implies with (\ref{FForm}) that
\be
F_{A B C} \ = \ F_{[A B C]} \ .
\ee
Note that these tensors (in the following sometimes referred to as fluxes) are generally
$X$-dependent.

After using the described $GL(D)\times GL(D)$ gauge fixing (\ref{GaugeFixing}) in which the fluctuations are constrained to satisfy $h_{ab} = h_{\bar a \bar b} = 0$ and $h_{a \bar b} = - h_{\bar b a}$, it is easy to see that the flat metric ${\cal G}_{AB}$ is exactly quadratic in fluctuations
\begin{eqnarray}
{\cal G}_{a b} &=& \bar {\cal G}_{a b} + h_a{}^{\bar c}  h_{b \bar c}\;,  \label{GabExp}\\
{\cal G}_{\bar a \bar b} &=& \bar {\cal G}_{\bar a \bar b} + h^c{}_{\bar a} h_{c \bar b} \ .
\end{eqnarray}
As before, we adopt the convention that flat indices of the background and fluctuations are raised and lowered with $\bar {\cal G}_{A B}$.  The inverse flat metric receives an infinite expansion
\begin{eqnarray}
{\cal G}^{a b} &=& \bar {\cal G}^{a b} - h^{a \bar c} h^b{}_{\bar c}  + {\cal O}(h^4)\;, \\
{\cal G}^{\bar a \bar b} &=& \bar {\cal G}^{\bar a \bar b} - h_c{}^{\bar a} h^{c \bar b} + {\cal O}(h^4) \ ,
\end{eqnarray}
but to cubic order the terms quartic and higher are not needed. The flat derivatives now take the form
\be
E_{a} \ = \ D_a - h_{a}{}^{\bar b} D_{\bar b} \ , \ \ \ \ \ E_{\bar a} \ = \ D_{\bar a} + h^b{}_{\bar a} D_b \ ,
\ee
which are exact.
Moreover, we find
\begin{eqnarray}
{\cal F}_a &=&  F_a - h_a{}^{\bar b} F_{\bar b}  - D_{\bar b} h_a{}^{\bar b} - 2 D_a d' + 2 h_a{}^{\bar b} D_{\bar b} d' \;,  \\
\Omega_{ab\bar c} &=&  F_{a b \bar c} + h^{d}{}_{\bar c} F_{a b d} - 2 D_{[a} h_{b] \bar c} + 2 h_{[a|}{}^{\bar d} D_{\bar d} h_{|b] \bar c} + D_{\bar c} h_{[a}{}^{\bar e} h_{b]\bar e}
- h_{\bar c}{}^d D_d h_{[a}{}^{\bar e} h_{b]\bar e} \\
&&  + 2 h_{[a}{}^{\bar d} F_{b]\bar d \bar c} + 2 h^d{}_{\bar c} h_{[a}{}^{\bar e} F_{b] \bar e d}
+ h_{a}{}^{\bar d} h_{b}{}^{\bar e} F_{\bar d \bar e \bar c} + h_{a}{}^{\bar d} h_{b}{}^{\bar e} h^{f}{}_{\bar c} F_{\bar d \bar e f} \;, \nonumber\\
\Omega_{[abc]} &=&  F_{a b c} - 3 h_{[a}{}^{\bar d} F_{b c] \bar d} + 3 D_{[a} h_b{}^{\bar d} h_{c] \bar d} - 3 h_{[a|}{}^{\bar d} D_{\bar d} h_{|b}{}^{\bar e} h_{c]\bar e} \label{OmegaExp}\\
&& + 3 h_{[a}{}^{\bar d} h_{b}{}^{\bar e} F_{c]\bar d \bar e} - h_{a}{}^{\bar d} h_b{}^{\bar e} h_{c}{}^{\bar f} F_{\bar d \bar e \bar f} \ . \nonumber
\end{eqnarray}

Finally, plugging the decompositions (\ref{GabExp}) -- (\ref{OmegaExp}) into the generalized Ricci scalar (\ref{FullGenRicci}) one obtains the expansion in fluctuations to cubic order. To present the result in a more readable form, we decompose it according to the powers of perturbations, \be
{\cal R} = {\cal R}_0 + {\cal R}_1 + {\cal R}_2 + {\cal R}_3 \ ,
\ee
where the label indicates the power counting in perturbations $h_{a \bar b}$ and $d'$. We now drop the prime on the dilaton, $d' \to d$, as it is clear from the context that it denotes a fluctuation. All in all, we find
\begin{eqnarray}
{\cal R}_0 &=& -4 {D}^{a}{{F}_{a}}\,  -2 \, {F}^{a} {F}_{a} + \frac{1}{3}\, {F}^{a b c} {F}_{a b c} +\, {F}^{a b \bar a} {F}_{a b \bar a} \ , \label{R0a}\\
{\cal R}_1 &=&  8\, {D}^{a}{{D}_{a}{d}\, }\,  + 4 {D}^{a}{{D}^{\bar a}{{h}_{a \bar a}}\, }\,  + 8\, {D}^{a}{d}\,  {F}_{a} + 4 {D}^{a}{{F}^{\bar a}}\,  {h}_{a \bar a} + 4 {D}^{a}{{h}_{a}\,^{\bar a}}\,  {F}_{\bar a} + 4 {D}^{\bar a}{{F}^{a}}\,  {h}_{a \bar a} \ \ \ \ \ \ \ \ \ \ \ \ \ \ \nonumber \\ && +\, 4 {D}^{\bar a}{{h}^{a}\,_{\bar a}}\,  {F}_{a} -4 {D}^{a}{{h}^{b \bar a}}\,  {F}_{a b \bar a} +4 {F}^{a} {F}^{\bar a} {h}_{a \bar a} + 4 {F}^{a b \bar a} {F}_{a \bar a}\,^{\bar b} {h}_{b \bar b} \ ,
\end{eqnarray}
\begin{eqnarray}
{\cal R}_2 &=& -8\, {D}^{a}{d}\,  {D}_{a}{d}\,  -8\, {D}^{a}{{D}^{\bar a}{d}\, }\,  {h}_{a \bar a} -8\, {D}^{a}{d}\,  {D}^{\bar a}{{h}_{a \bar a}}\,  -8\, {D}^{a}{{h}_{a}\,^{\bar a}}\,  {D}_{\bar a}{d}\,  -8\, {D}^{\bar a}{{D}^{a}{d}\, }\,  {h}_{a \bar a} \nonumber \\ && +\, 2\, {D}^{a}{{D}^{b}{{h}_{a}\,^{\bar a}}\, }\,  {h}_{b \bar a} + 2\, {D}^{a}{{D}^{b}{{h}_{b}\,^{\bar a}}\, }\,  {h}_{a \bar a} -8\, {D}^{a}{d}\,  {F}^{\bar a} {h}_{a \bar a} +2\, {D}^{a}{{h}_{a}\,^{\bar a}}\,  {D}^{b}{{h}_{b \bar a}}\, \nonumber \\ && +\, 2\, {D}^{a}{{h}^{b \bar a}}\,  {D}_{a}{{h}_{b \bar a}}\,  -4 {D}^{\bar a}{{D}^{\bar b}{{h}^{a}\,_{\bar b}}\, }\,  {h}_{a \bar a} -8\, {D}^{\bar a}{d}\,  {F}^{a} {h}_{a \bar a} -2\, {D}^{\bar a}{{h}^{a}\,_{\bar a}}\,  {D}^{\bar b}{{h}_{a \bar b}}\, \nonumber \\ &&  +\, 4 {D}^{a}{{F}^{b}}\,  {h}_{a}\,^{\bar a} {h}_{b \bar a} +4 {D}^{a}{{h}_{a}\,^{\bar a}}\,  {F}^{b} {h}_{b \bar a} +4 {D}^{a}{{h}^{b \bar a}}\,  {F}_{b} {h}_{a \bar a} -4 {D}^{\bar a}{{F}^{\bar b}}\,  {h}^{a}\,_{\bar a} {h}_{a \bar b} -4 {D}^{\bar a}{{h}^{a}\,_{\bar a}}\,  {F}^{\bar b} {h}_{a \bar b} \ \ \ \ \nonumber \\ && -\, 4 {D}^{\bar a}{{h}^{a \bar b}}\,  {F}_{\bar b} {h}_{a \bar a} -2\, {D}^{a}{{h}^{b \bar a}}\,  {F}_{a b}\,^{c} {h}_{c \bar a} -4 {D}^{a}{{h}^{b \bar a}}\,  {F}_{a \bar a}\,^{\bar b} {h}_{b \bar b} +4 {D}^{a}{{h}^{b \bar a}}\,  {F}_{b \bar a}\,^{\bar b} {h}_{a \bar b} \nonumber \\ && -\, 2\, {D}^{\bar a}{{h}^{a \bar b}}\,  {F}_{a \bar a}\,^{b} {h}_{b \bar b} +4 {D}^{\bar a}{{h}^{a \bar b}}\,  {F}_{a \bar b}\,^{b} {h}_{b \bar a} +2\, {F}^{a} {F}^{b} {h}_{a}\,^{\bar a} {h}_{b \bar a} -2\, {F}^{\bar a} {F}^{\bar b} {h}^{a}\,_{\bar a} {h}_{a \bar b} \nonumber \\&& -\, 2\, {F}^{a b c} {F}_{a}\,^{\bar a \bar b} {h}_{b \bar a} {h}_{c \bar b} +2\, {F}^{a b \bar a} {F}_{a \bar a}\,^{c} {h}_{b}\,^{\bar b} {h}_{c \bar b} +2\, {F}^{a b \bar a} {F}_{a}\,^{c \bar b} {h}_{b \bar b} {h}_{c \bar a} +2 \, {F}^{a b \bar a} {F}_{\bar a}\,^{\bar b \bar c} {h}_{a \bar b} {h}_{b \bar c} \ \ \ \ \nonumber \\ && +\, 2\, {F}^{a \bar a \bar b} {F}_{a \bar a}\,^{\bar c} {h}^{b}\,_{\bar b} {h}_{b \bar c} +2\, {F}^{a \bar a \bar b} {F}_{\bar a}\,^{b \bar c} {h}_{a \bar c} {h}_{b \bar b} \ ,
\end{eqnarray}
\begin{eqnarray}
{\cal R}_3 &=& 16\, {D}^{a}{d}\,  {D}^{\bar a}{d}\,  {h}_{a \bar a} -8\, {D}^{a}{{D}^{b}{d}\, }\,  {h}_{a}\,^{\bar a} {h}_{b \bar a} -8\, {D}^{a}{d}\,  {D}^{b}{{h}_{a}\,^{\bar a}}\,  {h}_{b \bar a} -8\, {D}^{a}{d}\,  {D}^{b}{{h}_{b}\,^{\bar a}}\,  {h}_{a \bar a} \nonumber \\ && +\, 8\, {D}^{\bar a}{{D}^{\bar b}{d}\, }\,  {h}^{a}\,_{\bar a} {h}_{a \bar b} + 8\, {D}^{\bar a}{d}\,  {D}^{\bar b}{{h}^{a}\,_{\bar a}}\,  {h}_{a \bar b} + 8\, {D}^{\bar a}{d}\,  {D}^{\bar b}{{h}^{a}\,_{\bar b}}\,  {h}_{a \bar a} -2\, {D}^{a}{{D}^{\bar a}{{h}_{a}\,^{\bar b}}\, }\,  {h}^{b}\,_{\bar a} {h}_{b \bar b}  \nonumber \\ && - \, 4\,  {D}^{a}{{D}^{\bar a}{{h}^{b}\,_{\bar a}}\, }\,  {h}_{a}\,^{\bar b} {h}_{b \bar b} -2\, {D}^{a}{{D}^{\bar a}{{h}^{b \bar b}}\, }\,  {h}_{a \bar b} {h}_{b \bar a} -8\, {D}^{a}{d}\,  {F}^{b} {h}_{a}\,^{\bar a} {h}_{b \bar a} -4 {D}^{a}{{h}_{a}\,^{\bar a}}\,  {D}^{\bar b}{{h}^{b}\,_{\bar b}}\,  {h}_{b \bar a}  \nonumber \\ &&  -\, 4 {D}^{a}{{h}^{b \bar a}}\,  {D}_{\bar a}{{h}_{a}\,^{\bar b}}\,  {h}_{b \bar b} -4 {D}^{a}{{h}^{b \bar a}}\,  {D}^{\bar b}{{h}_{b \bar b}}\,  {h}_{a \bar a} -4 {D}^{a}{{h}_{a}\,^{\bar a}}\,  {D}^{\bar b}{{h}^{b}\,_{\bar a}}\,  {h}_{b \bar b} -4 {D}^{a}{{h}^{b \bar a}}\,  {D}^{\bar b}{{h}_{b \bar a}}\,  {h}_{a \bar b}  \nonumber \\ &&  -\, 2 \, {D}^{\bar a}{{D}^{a}{{h}_{a}\,^{\bar b}}\, }\,  {h}^{b}\,_{\bar a} {h}_{b \bar b} -2\, {D}^{\bar a}{{D}^{a}{{h}^{b \bar b}}\, }\,  {h}_{a \bar b} {h}_{b \bar a} +8\, {D}^{\bar a}{d}\,  {F}^{\bar b} {h}^{a}\,_{\bar a} {h}_{a \bar b} -4 {D}^{a}{{F}^{\bar a}}\,  {h}_{a}\,^{\bar b} {h}^{b}\,_{\bar a} {h}_{b \bar b}  \nonumber \\ && -\, 4 {D}^{a}{{h}_{a}\,^{\bar a}}\,  {F}^{\bar b} {h}^{b}\,_{\bar a} {h}_{b \bar b} -4 {D}^{a}{{h}^{b \bar a}}\,  {F}_{\bar a} {h}_{a}\,^{\bar b} {h}_{b \bar b} -4 {D}^{a}{{h}^{b \bar a}}\,  {F}^{\bar b} {h}_{a \bar a} {h}_{b \bar b} -4 {D}^{\bar a}{{F}^{a}}\,  {h}_{a}\,^{\bar b} {h}^{b}\,_{\bar a} {h}_{b \bar b}  \nonumber \\ && -4 {D}^{\bar a}{{h}^{a \bar b}}\,  {F}_{a} {h}^{b}\,_{\bar a} {h}_{b \bar b} -4 {D}^{\bar a}{{h}^{a}\,_{\bar a}}\,  {F}^{b} {h}_{a}\,^{\bar b} {h}_{b \bar b} -4 {D}^{\bar a}{{h}^{a \bar b}}\,  {F}^{b} {h}_{a \bar a} {h}_{b \bar b} +2 \, {D}^{a}{{h}^{b \bar a}}\,  {F}_{a b}\,^{\bar b} {h}^{c}\,_{\bar a} {h}_{c \bar b}  \nonumber \\ && -\, 2\, {D}^{a}{{h}^{b \bar a}}\,  {F}_{a}\,^{c \bar b} {h}_{b \bar b} {h}_{c \bar a} +4 {D}^{a}{{h}^{b \bar a}}\,  {F}_{b}\,^{c \bar b} {h}_{a \bar b} {h}_{c \bar a} -4 {D}^{a}{{h}^{b \bar a}}\,  {F}_{a \bar a}\,^{c} {h}_{b}\,^{\bar b} {h}_{c \bar b} + 4 {D}^{a}{{h}^{b \bar a}}\,  {F}_{b \bar a}\,^{c} {h}_{a}\,^{\bar b} {h}_{c \bar b} \nonumber \\ &&  -\, 4\,  {D}^{a}{{h}^{b \bar a}}\,  {F}_{\bar a}\,^{\bar b \bar c} {h}_{a \bar b} {h}_{b \bar c} -4 {D}^{\bar a}{{h}^{a \bar b}}\,  {F}_{a}\,^{b c} {h}_{b \bar a} {h}_{c \bar b} -4 {D}^{\bar a}{{h}^{a \bar b}}\,  {F}_{a \bar b}\,^{\bar c} {h}^{b}\,_{\bar a} {h}_{b \bar c} -4 {D}^{\bar a}{{h}^{a \bar b}}\,  {F}_{\bar b}\,^{b \bar c} {h}_{a \bar c} {h}_{b \bar a}  \nonumber \\ && +\, 2\, {D}^{\bar a}{{h}^{a \bar b}}\,  {F}_{a \bar a}\,^{\bar c} {h}^{b}\,_{\bar b} {h}_{b \bar c} + 2 \, {D}^{\bar a}{{h}^{a \bar b}}\,  {F}_{\bar a}\,^{b \bar c} {h}_{a \bar c} {h}_{b \bar b} -4 {F}^{a} {F}^{\bar a} {h}_{a}\,^{\bar b} {h}^{b}\,_{\bar a} {h}_{b \bar b} +4 {F}^{a b c} {F}_{a}\,^{d \bar a} {h}_{b \bar a} {h}_{c}\,^{\bar b} {h}_{d \bar b}  \nonumber \\ &&  +\, \frac{4}{3}\, {F}^{a b c} {F}^{\bar a \bar b \bar c} {h}_{a \bar a} {h}_{b \bar b} {h}_{c \bar c} -4 {F}^{a b \bar a} {F}_{a \bar a}\,^{\bar b} {h}_{b}\,^{\bar c} {h}^{c}\,_{\bar b} {h}_{c \bar c} + 4 {F}^{a b \bar a} {F}_{a}\,^{\bar b \bar c} {h}_{b \bar b} {h}^{c}\,_{\bar a} {h}_{c \bar c}  \nonumber \\ &&  -\, 4\,  {F}^{a b \bar a} {F}_{\bar a}\,^{c \bar b} {h}_{a \bar b} {h}_{b}\,^{\bar c} {h}_{c \bar c} + 4 {F}^{a b \bar a} {F}^{c \bar b \bar c} {h}_{a \bar b} {h}_{b \bar c} {h}_{c \bar a} + 4 {F}^{a \bar a \bar b} {F}_{\bar a}\,^{\bar c \bar d} {h}_{a \bar c} {h}^{b}\,_{\bar b} {h}_{b \bar d} \ .\label{R3a}
\end{eqnarray}
It is a long but straightforward computation\footnote{We have benefited from \textsc{Cadabra} \cite{Peeters:2007wn} in this and other computations in this paper. } to verify, as consistency check, that the expanded Ricci scalar indeed transforms as a scalar with respect to the gauge transformations of the fluctuations (\ref{dilatongauge}) and (\ref{DELTAhSTEP}),
\be
\delta {\cal R} \ = \ \xi^A D_A {\cal R} \ , \label{Rscalar}
\ee
to cubic order, provided the strong constraint holds and making repeated use of the
relations  (\ref{FIdentities})--(\ref{LastFIdentities}).

We now want to rewrite the background expanded Ricci scalar in a background covariant form. To this end, we have
to use the full background Riemann tensor
\be
\bar{\cal R}_{ABCD} \ = \ \tfrac 1 2 \left(\bar R_{ABCD} + \bar R_{CD AB} - \bar \omega_{EAB} \bar \omega^E{}_{C D}\right) \ ,\label{BndGenRiemann}
\ee
with
\be
\bar R_{A B C D} = D_A \bar \omega_{BCD} - D_B \bar \omega_{ACD} + \bar \omega_{AC}{}^E \bar \omega_{B E D}- \bar \omega_{BC}{}^E \bar \omega_{AED} - F_{A B}{}^E \bar \omega_{ECD} \ .
\ee
The procedure is straightforward: we simply replace flat partial derivatives $D_A$ by background covariant derivatives $\bar \nabla_A$ plus the compensating background generalized spin connection terms. After this, most terms with spin connections cancel among each other, and the rest combines into non-vanishing components of the background generalized Riemann tensor (\ref{BndGenRiemann}).
We note that this requires using projections/contractions of the generalized Riemann tensor that contain
undetermined connection components, but as for the gauge transformations, they drop out of the full action.
After some algebra we find for (\ref{R0a})--(\ref{R3a})
\begin{eqnarray}
{\cal R}_0 &=& -2\, {\bar {\cal R}}^{a b}\,_{b a} \ , \\
{\cal R}_1 &=& 8\, {\bar \nabla}^{a}{{\bar \nabla}_{a}{d}\, }\,  + 4 {\bar \nabla}^{a}{{\bar \nabla}^{\bar a}{{h}_{a \bar a}}\, }\,  + 8 \, {\bar {\cal R}}^{a b \bar a}\,_{a} {h}_{b \bar a} \ ,\\
{\cal R}_2 &=& -\, 8\, {\bar \nabla}^{a}{d}\,  {\bar \nabla}_{a}{d}\,  - 8\, {\bar \nabla}^{a}{{h}_{a}\,^{\bar a}}\,  {\bar \nabla}_{\bar a}{d}\,  - 8\, {\bar \nabla}^{\bar a}{{h}_{a \bar a}}\,  {\bar \nabla}^{a}{d}\,  - 8\, {h}_{a \bar a} {\bar \nabla}^{a}{{\bar \nabla}^{\bar a}{d}\, }\,  - 8\, {h}_{a \bar a} {\bar \nabla}^{\bar a}{{\bar \nabla}^{a}{d}\, }\, \\ && +\, 2 \, {\bar \nabla}^{a}{{h}_{a}\,^{\bar a}}\,  {\bar \nabla}^{b}{{h}_{b \bar a}}\, + 2\, {\bar \nabla}^{a}{{h}^{b \bar a}}\,  {\bar \nabla}_{a}{{h}_{b \bar a}}\,  + 4 {h}_{a \bar a} {\bar \nabla}^{b}{{\bar \nabla}^{a}{{h}_{b}\,^{\bar a}}\, }\,  - 2\, {\bar \nabla}^{\bar a}{{h}^{a}\,_{\bar a}}\,  {\bar \nabla}^{\bar b}{{h}_{a \bar b}}\, \nonumber \\ && -\, 4 {h}_{a \bar a} {\bar \nabla}^{\bar a}{{\bar \nabla}^{\bar b}{{h}^{a}\,_{\bar b}}\, }\, + 4 {h}_{a}\,^{\bar a} {h}_{b \bar a} {\bar {\cal R}}^{c a b}\,_{c} + 4 h_{a \bar a} h_{b \bar b} \bar {\cal R}^{a b \bar a \bar b} \ , \nonumber \\
{\cal R}_3 &=& 16\, {h}_{a \bar a} {\bar \nabla}^{a}{d}\,  {\bar \nabla}^{\bar a}{d}\,  - 8\, {h}_{a \bar a} {\bar \nabla}^{b}{{h}_{b}\,^{\bar a}}\,  {\bar \nabla}^{a}{d}\,  - 8\, {h}_{a \bar a} {\bar \nabla}^{a}{{h}_{b}\,^{\bar a}}\,  {\bar \nabla}^{b}{d}\,  - 8\, {h}_{a}\,^{\bar a} {h}_{b \bar a} {\bar \nabla}^{a}{{\bar \nabla}^{b}{d}\, }\, \\ && + \, 8\, {h}_{a \bar a} {\bar \nabla}^{\bar b}{{h}^{a}\,_{\bar b}}\,  {\bar \nabla}^{\bar a}{d}\, + 8\, {h}_{a \bar a} {\bar \nabla}^{\bar a}{{h}^{a}\,_{\bar b}}\,  {\bar \nabla}^{\bar b}{d}\,  + 8\, {h}^{a}\,_{\bar a} {h}_{a \bar b} {\bar \nabla}^{\bar a}{{\bar \nabla}^{\bar b}{d}\, }\, \nonumber \\ && -\, 4 {h}_{a \bar a} {\bar \nabla}^{a}{{h}^{b \bar a}}\,  {\bar \nabla}^{\bar b}{{h}_{b \bar b}}\, - 4 {h}_{a \bar a} {\bar \nabla}^{a}{{h}^{b \bar b}}\,  {\bar \nabla}^{\bar a}{{h}_{b \bar b}}\,  - 4 {h}_{a \bar a} {\bar \nabla}^{b}{{h}_{b}\,^{\bar a}}\,  {\bar \nabla}^{\bar b}{{h}^{a}\,_{\bar b}}\, - 4 {h}_{a \bar a} {\bar \nabla}^{b}{{h}_{b}\,^{\bar b}}\,  {\bar \nabla}^{\bar a}{{h}^{a}\,_{\bar b}}\, \nonumber \\ && -\, 4 {h}_{a \bar a} {\bar \nabla}^{b}{{h}^{a \bar b}}\,  {\bar \nabla}_{\bar b}{{h}_{b}\,^{\bar a}}\, - 4 {h}_{a}\,^{\bar a} {h}_{b \bar a} {\bar \nabla}^{a}{{\bar \nabla}^{\bar b}{{h}^{b}\,_{\bar b}}\, }\,  - 2\, {h}_{a \bar a} {h}_{b \bar b} {\bar \nabla}^{a}{{\bar \nabla}^{\bar b}{{h}^{b \bar a}}\, }\,  - 2\, {h}^{a}\,_{\bar a} {h}_{a \bar b} {\bar \nabla}^{b}{{\bar \nabla}^{\bar a}{{h}_{b}\,^{\bar b}}\, }\, \nonumber \\ && - \, 2\, {h}_{a \bar a} {h}_{b \bar b} {\bar \nabla}^{\bar b}{{\bar \nabla}^{a}{{h}^{b \bar a}}\, }\,  - 2 \, {h}^{a}\,_{\bar a} {h}_{a \bar b} {\bar \nabla}^{\bar a}{{\bar \nabla}^{b}{{h}_{b}\,^{\bar b}}\, }\,  - 4 {h}_{a}\,^{\bar a} {h}_{b \bar b} {h}_{c \bar a} {\bar {\cal R}}^{a b \bar b c} - 4 {h}_{a}\,^{\bar a} {h}^{b}\,_{\bar b} {h}_{b \bar a} {\bar {\cal R}}^{c a \bar b}\,_{c} \ .\nonumber
\end{eqnarray}

Notice that ${\cal R}_0$ is proportional to the background generalized Ricci scalar, and $\bar {\cal R}_{c a \bar b}{}^c$ is the background generalized Ricci tensor. The last term in ${\cal R}_2$  turns out to vanish due to algebraic
Bianchi identities satisfied by the Riemann tensor
 \cite{Hohm:2012mf,Hohm:2011si}, so we will neglect if from now on.

\subsection{Expansion of the DFT action}

Having at hand the background field expansion of the generalized Ricci scalar in a background covariant form, we are now ready to compute the action to cubic order. We first write it in the following form
\be
S = \int dX e^{-2 \bar d} \left( {\cal L} + e^{-2 d} \lambda\right) \ ,
\ee
where
\be
{\cal L} = {\cal R}_0 +  {\cal R}_1 - 2 d  {\cal R}_0 +   {\cal R}_2 - 2 d  {\cal R}_1 + 2 d^2  {\cal R}_{0} +  {\cal R}_3 - 2 d  {\cal R}_2 + 2 d^2  {\cal R}_1 - \frac 4 3 d^3   {\cal R}_{0} \ ,
\ee
decomposed as a sum of ${\cal L}_{d_0 , h_0}$ where $(d_0 , h_0)$ represent the powers of $d$ and $h_{a \bar b}$ respectively. For this computation it is instrumental to recall that
\be
e^{-2 \bar d}\, \bar \nabla_a V^a \ = \  {\rm t. d.} \ , \ \ \ \ \ e^{-2 \bar d}\, \bar \nabla_{\bar a} V^{\bar a} \ = \  {\rm t. d.} \ ,
\ee
where ``t.d.'' stands for ``total derivative'', so any term of this form can be dropped. With this in mind, we find
\begin{eqnarray}
{\cal L}_{0 , 0} &=& -\, 2\, {\bar {\cal R}}^{a b}\,_{b a} \ , \\
{\cal L}_{1 , 0} &=&  4\, d {\bar {\cal R}}^{a b}\,_{b a} \ , \\
{\cal L}_{0 , 1} &=& 8\, {\bar {\cal R}}^{a b \bar a}\,_{a} {h}_{b \bar a} \ ,\\
{\cal L}_{2 , 0} &=& -\, 8\, d {\bar \nabla}^{a}{{\bar \nabla}_{a}{d}\, }\,  - 4 {d}^{2} {\bar {\cal R}}^{a b}\,_{b a} \ ,\\
{\cal L}_{1 , 1} &=&  -\, 8\, d {\bar \nabla}^{a}{{\bar \nabla}^{\bar a}{{h}_{a \bar a}}\, }\,  - 16\, d {\bar {\cal R}}^{a b \bar a}\,_{a} {h}_{b \bar a} \ , \\
{\cal L}_{0 , 2} &=& -\, 2 \, {\bar \nabla}^{a}{h}_{a \bar a}\,  {\bar \nabla}_{b}{{h}^{b \bar a}}\, - 2 \,{{h}^{b \bar a}}\,  {\bar \nabla}^{a}{\bar \nabla}_{a}{{h}_{b \bar a}}\,   + 2 \, {\bar \nabla}^{\bar a}{{h}_{a \bar a}}\,  {\bar \nabla}_{\bar b}{{h}^{a \bar b}}\, \\ &&
 -\, 8 {h}_{a \bar a} {\bar \nabla}^{[a}{{\bar \nabla}^{b]}{{h}_{b}\,^{\bar a}}\, }\, + 4 {h}_{a}\,^{\bar a} {h}_{b \bar a} {\bar {\cal R}}^{c a b}\,_{c}  \ , \nonumber \\
{\cal L}_{3 , 0} &=&   8 \, {d}^{2} {\bar \nabla}^{a}{{\bar \nabla}_{a}{d}\, }\,  + \frac{8}{3}\, {d}^{3} {\bar {\cal R}}^{a b}\,_{b a} \ , \\
{\cal L}_{2 , 1} &=& 16\, d\, {\bar \nabla}^{\bar a}{{\bar \nabla}^{a}{d}\, }\, {h}_{a \bar a}\,  + 16 \, {d}^{2} {\bar {\cal R}}^{a b \bar a}\,_{a} {h}_{b \bar a} \ , \\
{\cal L}_{1 , 2} &=& -\, 4\,  d \left[{\bar \nabla}^{a}{{h}_{a}\,^{\bar a}}\,  {\bar \nabla}^{b}{{h}_{b \bar a}}\,  +  {\bar \nabla}^{a}{{h}^{b \bar a}}\,  {\bar \nabla}_{a}{{h}_{b \bar a}}\, -  {\bar \nabla}^{\bar a}{{h}^{a}\,_{\bar a}}\,  {\bar \nabla}^{\bar b}{{h}_{a \bar b}}\,   \right. \\ && \left. \ \ \ \ \ \ \ \ +\, 2\,  {h}_{a \bar a} \left( {\bar \nabla}^{b}{{\bar \nabla}^{a}{{h}_{b}\,^{\bar a}}\, }\, - {\bar \nabla}^{\bar a}{{\bar \nabla}^{\bar b}{{h}^{a}\,_{\bar b}}\, } \right)\,  + 2\,  {h}_{a}\,^{\bar a} {h}_{b \bar a} {\bar {\cal R}}^{c a b}\,_{c} \right] \ , \nonumber\\
{\cal L}_{0 , 3} &=& -\, 4\, {h}_{a \bar a} \left(\bar \nabla^a h^{b \bar b}\, \bar \nabla^{\bar a} h_{b \bar b} - \bar \nabla^a h_{b \bar b}\, \bar \nabla^{\bar b} h^{b \bar a} - \bar \nabla^b h^{a \bar b} \, \bar \nabla^{\bar a} h_{b \bar b} + h^{b \bar a}\, \bar \nabla^{[a} \bar \nabla^{\bar b]} h_{b \bar b}\, \right) \\ &&  -\, 4\, {h}_{a}\,^{\bar a} {h}_{b \bar b} {h}_{c \bar a} {\bar {\cal R}}^{a b \bar b c} - 4 {h}_{a}\,^{\bar a} {h}^{b}\,_{\bar b} {h}_{b \bar a} {\bar {\cal R}}^{c a \bar b}\,_{c}  \ .\nonumber
\end{eqnarray}
Recall that using the background field equations
\be
\bar {\cal R} + \lambda = - 2 \bar {\cal R}^{a b}{}_{b a} + \lambda = 0 \ , \ \ \ \ \bar {\cal R}_{a \bar b} = 2 \bar {\cal R}_{c a \bar b}{}^c = 0 \ ,
\ee
some terms in the action vanish, and one arrives at the background covariant cubic action
\begin{eqnarray}\label{finalcubicstuff}
S \!\!\!&=&\!\!\! \int dX\, e^{- 2 \bar d}\, \bigg[ -\, 2\, {\bar \nabla}^{a}{{h}_{a \bar a}}\,  {\bar \nabla}_{b}{{h}^{b \bar a}}\, -\, 2\, {\bar \nabla}^{a}{{\bar \nabla}_{a}{{h}^{b \bar a}}\, }\,  {h}_{b \bar a} +\, 2 \, {\bar \nabla}^{\bar a}{{h}_{a \bar a}}\,  {\bar \nabla}_{\bar b}{{h}^{a \bar b}}\,   \\ && \ \ \ \ \ \ \ \ \ \ \ \ \ \ \ \ -\, 8\, {\bar \nabla}^{[a}{{\bar \nabla}^{b]}{{h}_{b}\,^{\bar a}}\, }\,  {h}_{a \bar a} + 4\, {\bar {\cal R}}^{a b c}\,_{a} {h}_{b}\,^{\bar a} {h}_{c \bar a} - 8\, d {\bar \nabla}^{a}{{\bar \nabla}^{\bar a}{{h}_{a \bar a}}\, }\,  - 8 \, d {\bar \nabla}^{a}{{\bar \nabla}_{a}{d}\, }\,  \nonumber \\ && \ \ \ \ \ \ \ \ \ \ \ \ \ \ \ \  -\, 4\, h_{a\bar b} \left({\bar \nabla}^{a}{{h}^{b \bar a}}\,  {\bar \nabla}^{\bar b}{{h}_{b \bar a}}\,  - {\bar \nabla}^{a}{{h}_{b \bar a}}\,  {\bar \nabla}^{\bar a}{{h}^{b \bar b}}\,  - {\bar \nabla}^{b}{{h}^{a \bar a}}\,  {\bar \nabla}^{\bar b}{{h}_{b \bar a}}\, + \, {\bar \nabla}^{[a}{{\bar \nabla}^{\bar a]}{{h}^{b}\,_{\bar a}}\, }\, {h}_{b}{}^{ \bar b}\right) \nonumber \\ && \ \ \ \ \ \ \ \ \ \ \ \ \ \ \ \  -\, 4\, {\bar {\cal R}}^{a b \bar a c} {h}_{a}\,^{\bar b} {h}_{b \bar a} {h}_{c \bar b} - 8\, d\, {\bar {\cal R}}^{a b c}\,_{a}  {h}_{b}\,^{\bar a} {h}_{c \bar a}  -\, 4 \, d \left[ {\bar \nabla}^{a}{{h}_{a \bar a}}\,  {\bar \nabla}_{b}{{h}^{b \bar a}}\, \right. \nonumber \\ &&  \ \ \ \ \ \ \ \ \ \ \ \ \ \ \ \ \left. +  {\bar \nabla}^{a}{{h}^{b \bar a}}\,  {\bar \nabla}_{a}{{h}_{b \bar a}}\, -  {\bar \nabla}_{\bar a}{{h}^{a \bar a}}\,  {\bar \nabla}^{\bar b}{{h}_{a \bar b}}\,  + 2\, h_{a \bar b} \left( {\bar \nabla}^{b}{{\bar \nabla}^{a}{{h}_{b}\,^{\bar b}}\, }\,  - \,  {\bar \nabla}^{\bar b}{{\bar \nabla}^{\bar a}{{h}^{a}\,_{\bar a}}\, } \right)\right]  \nonumber \\ &&   \ \ \ \ \ \ \ \ \ \ \ \ \ \ \ \  +\, 16 \, d\, {\bar \nabla}^{\bar a}{{\bar \nabla}^{a}{d}\, }\,  {h}_{a \bar a} + 8 \, {d}^{2} {\bar \nabla}^{a}{{\bar \nabla}_{a}{d}\, }\,  \bigg] \,.  \nonumber
\end{eqnarray}
This is the final form of the complete, background covariant cubic action, which may be specialized
to an arbitrary background solution.\footnote{We have verified that the quadratic part of the action agrees, up to an overall normalization, with that given in
\cite{Ko:2015rha} upon identifying $\delta P_{AB}\rightarrow -h_{a\bar{b}}-h_{b\bar{a}}$.}
For applications it may be more convenient to have a form in which the action is written out explicitly,
separating background connections in covariant derivatives
and curvatures. We give such expressions, and equivalent ones
in which barred and unbarred indices are on the same footing, in Appendix A.

\section{WZW Backgrounds}

\subsection{Generalities on group manifolds}\label{WZWsection}

We start by briefly reviewing the relevant aspects of group manifolds G.
Let the Lie group G with Lie algebra $\mathfrak{g}$ have generators $t_a$, $a=1,\ldots, n={\rm dim}\,\frak{g}$, with Lie bracket and Cartan-Killing form
 \be
  \big[ t_a,t_b\big] \  = \ f_{ab}{}^{c} t_c\;, \qquad
  \kappa_{ab} \ = \  \langle t_{a},t_b\rangle \ \equiv \ -f_{ac}{}^{d} f_{bd}{}^{c}\;.
 \ee
The quadratic form is invariant under the adjoint action of $g\in{\rm G}$ on $\frak{g}$ defined by
 \be\label{adjoint}
  g\, t^a g^{-1} \ \equiv \ t^b g_{b}{}^{a} \;,
 \ee
where $g_{a}{}^{b}$ is the group representative of $g$ in the adjoint representation.
Invariance means that for any $X, Y\in \frak{g}$ we have
 \be\label{INV}
  \langle g\,Xg^{-1},g\,Yg^{-1}\rangle \ = \ \langle X,Y\rangle\;.
 \ee
We introduce a group-valued function $\gamma(x)\in {\rm G}$, depending on
the coordinates $x^i$, $i=1,\ldots, n$, of the group manifold. This
can be used to define a $\frak{g}$-valued right-invariant Maurer-Cartan form and the corresponding metric on G by
 \be\label{RMC}
  \partial_i\gamma\, \gamma^{-1}\ \equiv \ e_{i}{}^{a}t_{a}\;, \qquad
  g_{ij} \ \equiv \ \langle    \partial_i\gamma\, \gamma^{-1},   \partial_j\gamma \, \gamma^{-1}\rangle \ = \
  e_{i}{}^{a}e_{j}{}^{b}\kappa_{ab}\;.
 \ee
Thus, the Maurer-Cartan form $e_{i}{}^{a}$ can be viewed as a vielbein for the metric $g_{ij}$ on G,
and can be expressed as
 \be\label{RIMC}
  e_{i}{}^{a} \ = \ \langle    \partial_i\gamma\,\gamma^{-1}, t^a\rangle\;.
 \ee
Under the following group action by rigid $g_{L}, g_{R}\in {\rm G}$
 \be\label{LRaction}
  \gamma\,\rightarrow \, g_L \,\gamma\, g_{R} \;, \qquad
  \partial_i\gamma\,\gamma^{-1}\ \rightarrow\ g_L(\partial_i\gamma\,\gamma^{-1})g_L^{-1}\;,
 \ee
we see that $e_{i}{}^{a}$ is right-invariant, while it transforms under
$g_L$ (in the adjoint representation) as indicated by the index $a$.
Since the Cartan-Killing metric $\kappa_{ab}$ is G-invariant, it follows
that $g_{ij}$ is invariant under ${\rm G}_L\times {\rm G}_R$, the isometry group of G.
Similarly, we can define left-invariant Maurer-Cartan forms by
  \be\label{MCL}
  \gamma^{-1}\partial_i\gamma \ \equiv \ \bar{e}_{i}{}^{\bar{a}}t_{\bar{a}}\;, \qquad
  g_{ij} \ \equiv \ \langle    \gamma^{-1}\partial_i\gamma,   \gamma^{-1}\partial_j\gamma \rangle \ = \
  \bar{e}_{i}{}^{\bar{a}}\bar{e}_{j}{}^{\bar{b}}\kappa_{\bar{a}\bar{b}}\;,
 \ee
and thus
 \be\label{LIMC}
  \bar{e}_{i}{}^{\bar{a}} \ = \ \langle    \gamma^{-1}\partial_i\gamma, t^{\bar{a}}\rangle\;,
 \ee
where we introduced a notation of barred indices in order to indicate that
it transforms under (\ref{LRaction}) in the adjoint representation of G$_R$,
while it is invariant under G$_L$.\footnote{We emphasize that at this stage there is no
difference between unbarred and barred indices. Both refer to the same Lie algebra $\frak{g}$.
However, this notation will be convenient momentarily in the doubled formalism.}
The ${\rm G}_L\times {\rm G}_R$ invariant metric $g_{ij}$
is the same as that defined in terms of $e_{i}{}^{a}$, which follows from the fact that both `vielbeine'
agree up to a G-transformation given by $\gamma$ itself,
 \be\label{LRrel}
  e_{i}{}^{a} \ = \ (\gamma^{-1})_{\bar{a}}{}^{a}\, \bar{e}_{i}{}^{\bar{a}}\;, \qquad
  \bar{e}_{i}{}^{\bar{a}} \ = \ \gamma_{a}{}^{\bar{a}} \, e_{i}{}^{a}\;,
 \ee
where $\gamma_{a}{}^{\bar{b}}$ is the representative of $\gamma$ in the adjoint representation
according to (\ref{adjoint}). These relations are easily verified: starting from the right-hand side
of the first equation we compute
 \be
  \begin{split}
   (\gamma^{-1})_{\bar{a}}{}^{a}\, \bar{e}_{i}{}^{\bar{a}} \ &= \    (\gamma^{-1})_{\bar{a}}{}^{a}\,
    \langle    \gamma^{-1}\partial_i\gamma, t^{\bar{a}}\rangle \ = \
     \langle    \gamma^{-1}\partial_i\gamma, \gamma^{-1}t^{a}\gamma\rangle \\[0.5ex]
     \ &= \ \langle \partial_i\gamma\,\gamma^{-1},t^a\rangle \ = \ e_{i}{}^{a}\;,
  \end{split}
 \ee
where we used the adjoint action (\ref{adjoint}) and the invariance property (\ref{INV}).
The analogous computation proves the second equation above.
Further useful relations follow from (\ref{RMC}) by taking $\gamma$ to be a matrix in
the adjoint representation,
 \be
  (\partial_i\gamma\,\gamma^{-1})_b{}^{c} \ = \ \partial_i\gamma_{b}{}^{\bar{d}}(\gamma^{-1})_{\bar{d}}{}^{c}
  \ = \ e_{i}{}^{a}(t_a)_{b}{}^{c} \ = \ -e_{i}{}^{a}f_{ab}{}^{c}\;,
 \ee
from which we conclude  $D_a\gamma_{b}{}^{\bar{c}}=-f_{ab}{}^{c}\gamma_{c}{}^{\bar{c}}$
with the flattened derivative $D_{a}\equiv e_{a}{}^{i}\partial_i$.
The analogous computation can be performed for (\ref{MCL}),
and one finds in total
  \be\label{gammaderrel}
   D_a\gamma_{b}{}^{\bar{c}} \ = \ -f_{ab}{}^{c}\,\gamma_{c}{}^{\bar{c}}\;, \qquad
   D_{\bar{a}}(\gamma^{-1})_{\bar{b}}{}^{c} \ = \ f_{\bar{a}\bar{b}}{}^{\bar{c}}\,(\gamma^{-1})_{\bar{c}}{}^{c}\;.
  \ee

In the following we need identities for the derivatives of the Maurer-Cartan forms.
We compute from (\ref{RIMC})
 \be\label{curlRF}
  \begin{split}
   \partial_i e_{j}{}^{a}- \partial_j e_{i}{}^{a} \ &= \ 2\,\big\langle -\partial_{[j}\gamma\,\gamma^{-1}
   \partial_{i]}\gamma\,\gamma^{-1},
   t^a\big\rangle \\[0.5ex]
   \ &= \ \big\langle [\partial_i\gamma\,\gamma^{-1},\partial_j\gamma\,\gamma^{-1}],t^a\big\rangle\\[0.5ex]
   \ &= \  \big\langle e_{i}{}^{b}e_{j}{}^{c}f_{bc}{}^{d}t_d,t^a\big\rangle\\[0.5ex]
   \ &= \ e_{i}{}^{b}e_{j}{}^{c}f_{bc}{}^{a}\;.
  \end{split}
 \ee
Comparing with the standard torsion constraint that determines the Levi-Civita spin-connection
we infer that the latter is determined to be (in flattened indices) $\omega^{\rm L}_{abc} = \tfrac{1}{2}f_{abc}$.
Another useful relation is obtained by contracting the above equation with $e_{a}{}^{j}$,
 \be\label{beincond}
  e_{a}{}^{j}\big(\partial_ie_{j}{}^{a}-\partial_j e_{i}{}^{a}\big) \ = \ 0\;,
 \ee
using the unimodularity condition
$f_{ba}{}^{a}=0$, which holds for any Lie algebra with non-degenerate
Cartan-Killing metric, as assumed here.
The analogous computation for the left-invariant form shows
 \be
  \partial_i\bar{e}_{j}{}^{\bar{a}}-\partial_j\bar{e}_{i}{}^{\bar{a}} \ = \ -
  \bar{e}_{i}{}^{\bar{b}} \bar{e}_{j}{}^{\bar{c}} f_{\bar{b}\bar{c}}{}^{\bar{a}}\;.
 \ee
We note that this differs from the result (\ref{curlRF}) for the right-invariant form by a global sign.
Consequently, the spin connection has the opposite sign, so that we have shown in total
that the connection components associated to the right- and left-invariant forms are, respectively,
 \be\label{Gspins}
   \omega^{\rm L}_{abc} \ = \  \tfrac{1}{2} f_{abc}\;, \qquad
   \bar{\omega}^{ \rm L }_{\bar{a}\bar{b}\bar{c}} \ = \ -\tfrac{1}{2} f_{\bar{a}\bar{b}\bar{c}}\;,
 \ee
where we indicated by the superscript L that these are the conventional
Levi-Civita spin connections, as opposed to the generalized connections of DFT.

\subsection{Frame field for WZW background}

We are now ready to define the generalized (background) frame field of DFT for WZW backgrounds,
which has also been given in  \cite{Blumenhagen:2015zma}.\footnote{See
also \cite{Schulz:2011ye} for earlier results on WZW models and doubled geometries.}
It can naturally be expressed in terms of the left- and right-invariant Maurer-Cartan forms,
 \be\label{BackgroundFrame}
  \bar{E}_{A}{}^{M} \ = \  \begin{pmatrix}  e_{ia} +B_{ij}e_{a}{}^{j} & e_{a}{}^{i} \\[0.5ex]
  -\bar{e}_{i\bar{a}}+B_{ij}\bar{e}_{\bar{a}}{}^{j} & \bar{e}_{\bar{a}}{}^{i} \end{pmatrix}\;,
 \ee
where the flat indices are raised and lowered with the Cartan-Killing metric $\kappa$.
Moreover, $B_{ij}$ is a two-form
whose field strength $H={\rm d}B$
in flat indices is given by\footnote{Both forms are equivalent. This follows from (\ref{LRrel})
since the conversion of barred to unbarred indices is governed by the G-valued
matrix $\gamma_{a}{}^{\bar{a}}$, under which the structure constants are invariant.}
 \be\label{HFORM}
  H_{abc} \ = \ -f_{abc}\;, \qquad H_{\bar{a}\bar{b}\bar{c}} \ = \ -f_{\bar{a}\bar{b}\bar{c}}\;.
 \ee
Such a two-form is locally guaranteed to exist up to gauge transformations, because the Bianchi identity
is satisfied:
 \be
  4\,\partial_{[i}H_{jkl]} \ = \ -12\, (\partial_{[i} e_{j}{}^{a})e_k{}^{b}e_{l]}{}^{c} f_{abc} \ = \
  -6\,e_{[i}{}^{d} e_{j}{}^{e} e_{k}{}^{b} e_{l]}{}^{c}\,f_{de}{}^{a}f_{abc} \ = \ 0\;,
 \ee
using (\ref{curlRF}) and the Jacobi identity.
By a straightforward computation one may confirm that
the above frame field leads to the background tangent space metric
 \be\label{BackgroundG}
   \bar{\cal G}_{AB} \ = \ \begin{pmatrix}  2\,\kappa_{ab} & 0 \\[0.5ex]
  0 & -2\,\kappa_{\bar{a}\bar{b}}  \end{pmatrix}\;,
 \ee
thus satisfying the constraint  (\ref{calGconstr}).
Due to the relative factors of $\pm 2$ in here one has to be careful when
translating index contractions with $\bar{\cal G}$ to index contractions with the Killing metric.
Moreover, it is evident that we have effectively  fixed a gauge for the $GL(D)\times GL(D)$
symmetry, which is reduced to a global ${\rm G}_L\times {\rm G}_R$,
and for which the flat background metric is constant.\footnote{While the metric $\bar{\cal G}_{AB}$
is even invariant under local ${\rm G}_L\times {\rm G}_R$ transformations, the perturbative action
is only invariant under the global subgroup.}

Let us now determine the background  $F_{ABC}$ defined in (\ref{FDef}).
Owing to the relative sign in (\ref{Gspins}) and using (\ref{HFORM}) one finds by direct computation
 \be\label{offzero}
  F_{\bar{a}bc} \ = \ 0\;, \qquad F_{a\bar{b}\bar{c}} \ = \ 0\;.
 \ee
In order to prove this one has to make repeated use of the relations (\ref{LRrel}) between
$e_{i}{}^{a}$ and $\bar{e}_{i}{}^{\bar{a}}$ and the invariance of the structure constants
under the action of $\gamma_{a}{}^{\bar{a}}$.
On the other hand, for the diagonal components, the spin connection and $H$ contributions combine,
 \be\label{structurconst}
  F_{abc} \ = \ -\,2 f_{abc}\;, \qquad
  F_{\bar{a}\bar{b}\bar{c}}  \ = \ -2\, f_{\bar{a}\bar{b}\bar{c}}\;.
 \ee
We recall that one has to carefully consider the index positions in order
to compare quantities. Here, for instance, on the $F_{ABC}$ on the left-hand sides indices
are raised and lowered with the background tangent space metric (\ref{BackgroundG}),
while on  the structure constant the adjoint indices are raised and lowered with the
Killing metric. Thus, taking into account the relative factors of $\pm 2$ in (\ref{BackgroundG}),
we infer
 \be\label{nonvanF}
  F_{ab}{}^{c} \ = \ -\,f_{ab}{}^{c}\;, \qquad
  F_{\bar{a}\bar{b}}{}^{\bar{c}} \ = \ \,f_{\bar{a}\bar{b}}{}^{\bar{c}}\;.
 \ee
In this form we can compare with eq.~(2.63) of \cite{Blumenhagen:2014gva}
and confirm in particular the relative sign difference.
These components determine the (traceless parts of the) background connections
according to  (\ref{connections}) in terms of the structure constants.

Next we determine the trace part of the background connections by computing
$F_A$ in  (\ref{BackFA}). We have to use that the background is independent of $\tilde{x}$
and that the scalar dilaton $\phi$ vanishes in the background, implying $e^{-2\bar{d}}=\sqrt{g}$.
We then compute
 \be
 \begin{split}
  F_i \ &\equiv \ e_{i}{}^{a} F_{a} \ = \ e_{i}{}^{a}\partial_je_a{}^{j}-2\,\partial_i\bar{d}
  \ = \ e_{i}{}^{a}\partial_je_a{}^{j} + e_a{}^{j} \partial_i e_{j}{}^{a}\\[0.5ex]
  \ &= \ e_{a}{}^{j}(\partial_ie_{j}{}^{a}-\partial_j e_{i}{}^{a}) \ = \ 0 \;,
 \end{split}
 \ee
where we used (\ref{beincond}) in the last step.
This implies $F_{a}=0$. An analogous computation shows $F_{\bar{a}}=0$,
so that we have proven in total
 \be\label{FAiszero}
  F_{A} \ = \ 0\qquad \Leftrightarrow \qquad \bar{E}_{M}{}^{A}F_A \ = \
  -D_{A}\bar{E}^A{}_{M}-2\,\partial_M\bar{d} \ = \ 0\;,
 \ee
where we recorded an equivalent form of this statement, which
is sometimes more useful. Thus, the trace part of the background spin connection
vanishes, $\bar{\omega}_A=0$.

We now can read off from (\ref{Curvatures}) the background generalized Ricci tensor
and curvature scalar in order to confirm that the WZW background solves the
field equations. With (\ref{structurconst}) we infer
 \be
 \begin{split}
  \bar{\cal R} \ \; &= \ \;  \tfrac{1}{3}\,F^{abc} F_{abc} \ = \ \tfrac{1}{3}\,4\,\tfrac{1}{8}\,f^{abc} f_{abc}
  \ = \ \tfrac{1}{6}\,{\rm dim}\,\frak{g}\;, \\[0.5ex]
  \bar{\cal R}_{a\bar{b}} \ &=  \ 0\;,
 \end{split}
 \ee
where we recalled in the first line the relative factors of $\pm \frac{1}{2}$ originating
from (\ref{BackgroundG}) that yield an additional $\tfrac{1}{2^3}$ when converting
a contraction with $\bar{\cal G}^{ab}$ to a contraction with $\kappa^{ab}$ as
employed in the last step. We conclude that the DFT field equations are satisfied
provided we choose the cosmological parameter to be $\lambda=-\tfrac{1}{6}\,{\rm dim}\,\frak{g}$,
which corresponds to string theory in a non-critical dimension.\footnote{Here we have suppressed possible
non-compact dimensions. For the inclusion of a Minkowski space factor and the corresponding analysis of the
critical dimension see Appendix B.}

We conclude this subsection by writing the strong constraint for the WZW backgrounds.
Since by (\ref{FAiszero}) we have $F_A=0$, it follows from (\ref{FIdentities}) that
 \be\label{flattConstr2}
   D^{A}D_{A} \ = \ D^aD_a+D^{\bar{a}}D_{\bar{a}} \ = \ 0\;,
 \ee
holds in general, acting on arbitrary fields and their products.
This constraint takes the same form as that given in \cite{Blumenhagen:2014gva},
again confirming the consistency of the string field theory and the DFT computation.

\subsection{Gauge algebra}
We now discuss the perturbative gauge structure for the WZW case.
The gauge transformations can be read off from (\ref{backcovgauge}),
using that by (\ref{connections}) and (\ref{offzero}) the only non-vanishing (totally antisymmetric)
background connections are
 \be
  \bar{\omega}_{ab}{}^{c} \ = \ -\tfrac{1}{3}F_{ab}{}^{c}\;, \qquad
  \bar{\omega}_{\bar{a}\bar{b}}{}^{\bar{c}} \ = \ -\tfrac{1}{3}F_{\bar{a}\bar{b}}{}^{\bar{c}}\;.
 \ee
Alternatively, one may use (\ref{DELTAhSTEP}) directly in order to arrive at the same result,
which reads
 \be\label{WZWgauge}
  \begin{split}\;
  \delta h_{a\bar{b}} \ = \ \;
   &D_{a}\xi_{\bar{b}}-D_{\bar{b}}\xi_{a}\\[0.5ex]
   &+\xi^{C}D_{C}h_{a\bar{b}}+\big(D_{a}\xi^{c}-D^{c}\xi_{a}\big)h_{c\bar{b}}
   +\big(D_{\bar{b}}\xi^{\bar{c}}-D^{\bar{c}}\xi_{\bar{b}}\big)h_{a\bar{c}} \\[0.5ex]
   &+h_{a\bar{d}}\big(D^{c}\xi^{\bar{d}}-D^{\bar{d}}\xi^{c}\big) h_{c\bar{b}} \\[0.5ex]
   &-\xi^c \,F_{ca}{}^{d} \,h_{d\bar{b}} - \xi^{\bar{c}}\, F_{\bar{c}\bar{b}}{}^{\bar{d}} \, h_{a\bar{d}}\;.
  \end{split}
 \ee
The terms in the first three lines take the same form as the gauge transformations on flat space,
except that the $D_A$ depend on the now $x$-dependent background, while on flat space
they depend on the constant background.  The terms in the fourth line encode the
deformation due to the structure constants $F$. Note that the $F$ enter the terms linear in $h$
but not the quadratic terms. For the cubic theory only the terms linear in $h$ are relevant.
Finally, the gauge transformations (\ref{dilatongauge}) of the dilaton fluctuation
reduce to
 \be
  \delta_{\xi}d' \ = \ \xi^A D_A d'-\tfrac{1}{2}D_A\xi^A\;,
 \ee
where we used $F_A=0$.

The above gauge transformations close according to the algebra (\ref{deformedbracket}),
which in the present case reduces to
 \be
  \begin{split}
   \xi_{12}^a \ &= \ 2\,\xi_{[2}^{B}D_{B}\xi_{1]}^a
   -\xi_{[2 B}D^a \xi_{1]}^B   \, + \, F^{a}{}_{bc}\,\xi_2^b\, \xi_1^c\;, \\[0.5ex]
   \xi_{12}^{\bar{a}} \ &= \ 2\,\xi_{[2}^{B}D_{B}\xi_{1]}^{\bar{a}} -\xi_{[2 B}D^{\bar{a}} \xi_{1]}^B
   \, + \, F^{\bar{a}}{}_{\bar{b}\bar{c}}\,\xi_2^{\bar{b}}\, \xi_1^{\bar{c}}\;.
 \end{split}
 \ee
Closure modulo the constraint (\ref{flattConstr}) and the Jacobi identities, $F_{[ab}{}^{d} F_{c]d}{}^{e}=0$, etc.,
may be verified directly, for which
one has to use the commutators implied by (\ref{DCommu}),
 \be\label{WZWFalgebra}
  \big[D_{a},D_{b}\big] \ = \ F_{ab}{}^{c}D_c\;, \qquad
  \big[D_{\bar{a}},D_{\bar{b}}\big] \ = \ F_{\bar{a}\bar{b}}{}^{\bar{c}}D_{\bar{c}}\;, \qquad
  \big[D_{a},D_{\bar{b}}\big] \ = \ 0\;.
 \ee
The gauge algebra takes the same form  as the original C-bracket written in flattened indices,
except for the `deformation' by the structure constants $F^a{}_{bc}$ and $F^{\bar{a}}{}_{\bar{b}\bar{c}}$.
Of course, it should be emphasized that this is not a real deformation as the
above algebra originates from the C-bracket by simply flattening the indices with the background
frame field, which is also implicit in the derivatives $D_A$.

\subsection{Cubic action for WZW backgrounds}
We can finally give the cubic action around a WZW background by specializing
the cubic action given in sec.~3 to the background frame (\ref{BackgroundFrame}),
for which we recall that the only non-vanishing flux components are
$F_{abc}$ and $F_{\bar a \bar b \bar c}$. Using this simplification and the explicit expressions (\ref{explicitA1})
and (\ref{explicitA2}) in Appendix A, leads to the following action
\begin{eqnarray}
S &=& \int dX\, e^{-2 \bar d}\, \bigg[ \, -2\, h_{a \bar b}\,\Box \, h^{a \bar b} + 2 D^{\bar a}h_{b \bar a} \, D_{\bar c} h^{b \bar c} -2 D^c h_{c \bar a} \, D_d h^{d \bar a} - 8 \, d\, D^a D^{\bar b} h_{a \bar b} - 8 \, d\, \Box\, d
\nonumber \\ && \ \ \ \ \ \ \ \ \ \ \ \ \ \ \ \ \  -\, 4\, h_{a\bar b} \left(D^a h_{c \bar d} D^{\bar b}h^{c \bar d} - D^a h_{c \bar d} D^{\bar d}h^{c \bar b} - D^c h^{a \bar d} D^{\bar b} h_{c \bar d}\right)  \\ &&\ \ \ \ \ \ \ \ \ \ \ \ \ \ \ \ \  -\, 4 \, d \left(D^b h_{b \bar a} D_dh^{d \bar a} - D^{\bar a} h_{b \bar a} D_{\bar c}h^{b \bar c} + \tfrac 1 2 D^c h^{d \bar a} D_c h_{d \bar a} - \tfrac 1 2 D^{\bar c} h^{d \bar a}D_{\bar c} h_{d \bar a} \right. \nonumber \\ && \ \ \ \ \ \ \ \ \ \ \ \ \ \ \ \ \ \ \  \left. \qquad \;\;
+\, 2 h_{a \bar b}\big(D^a D_c h^{c \bar b} - D^{\bar b}D_{\bar c} h^{a \bar c}\big)\right)  + 16\, h_{a \bar b}\,d\, D^a D^{\bar b} d\;  + 8 \, d^2\, \Box \, d\nonumber \\ && \ \ \ \ \ \ \ \ \  \ \ \ \ \ \ \ \  + \, 4\, h_{a \bar b} \left(F^{ac}{}_d\, D^{\bar e} h^{d \bar b}\, h_{c \bar e} + F^{\bar b \bar c}{}_{\bar d}\, D^e h^{a \bar d}\, h_{e \bar c}\right)
 + \tfrac 4 3 F^{a c e} \,F^{\bar b \bar d \bar f} \,h_{a \bar b} \, h_{c \bar d} \, h_{e \bar f} \bigg] \nonumber \;,
\end{eqnarray}
where we defined the box as
\be
\Box \ = \  D_a D^a \ = \  - D_{\bar a} D^{\bar a}\;.
\ee
This action is by construction invariant under gauge transformations
modulo the strong constraint, which by the results of the previous section can be written in the form
(\ref{flattConstr}), $D^AD_A=0$.

Let us now compare this action with that determined by Blumenhagen at.~al.~in
\cite{Blumenhagen:2014gva}. Comparing with eq.~(3.75) in that reference one
may confirm by inspection that they agree precisely under the following identifications
\be
h_{a \bar b} \ \to \, - \, \epsilon_{a \bar b} \ , \ \ \ \  \bar {\cal G}_{a b} \, \to \, 2\,\kappa_{a b} \ ,
\ \ \ \ \bar {\cal G}_{\bar{a}\bar{ b}} \, \to \, -2\,\kappa_{\bar{a}\bar{ b}}
\ \ \ \ e^{-2 \bar d}\, \to \, \sqrt{|H|}\;,
\ee
where $H$ is a (doubled) background metric introduced in \cite{Blumenhagen:2014gva} that is used to define
the background volume element, a role that in DFT is played by the background dilaton density.
In order to establish the above dictionary, in which one changes
the metric that is used to contract indices from $\bar{\cal G}_{AB}$ to $\kappa_{ab}$,
we have to specify which index positions are the basic ones, from which all others are obtained
by raising and lowering. These are
\be
h_{a \bar b} \ , \ \ \ F_{a b}{}^c \ , \ \ \ D_a\;.
\ee

Solving the strong constraint by setting $\tilde{\partial}^i=0$, we infer with the background
frame (\ref{BackgroundFrame}) that the flattened derivatives reduce to
 \be\label{strangeDs}
  D_{a} \ = \  e_{a}{}^{i}\partial_i\;, \qquad
  D_{\bar{a}} \ = \ \bar{e}_{\bar{a}}{}^{{i}}\partial_{{i}}\;.
 \ee
These agree with the derivative operators emerging naturally from string field theory on
WZW backgrounds, in which they correspond to the left- and right-invariant conserved currents of the worldsheet CFT
\cite{Blumenhagen:2014gva}. Let us emphasize that on
group manifolds such as $S^3\simeq SU(2)$ there are generally no winding modes
and hence there is no doubling of coordinates in the string field.\footnote{This is in contrast to
toroidal backgrounds, where the background is constant but the string field depends
on doubled coordinates.   We thank Barton Zwiebach
for discussions on this point.} Therefore, in DFT
one has to solve the strong constraint in order to compare with the string field theory action.

In order to compare  in more detail the above results with those in \cite{Blumenhagen:2014gva}
let us recall that the gauge invariance of the cubic action relies on the two relations
 \be\label{summREL}
  \big[D_A,D_B\big] \ = \ F_{AB}{}^{C}D_C\;, \qquad D^A D_A \ = \ 0\;.
 \ee
In the conventional DFT, the first relation holds for the $F_{ABC}$ as defined in (\ref{FDef})
modulo the strong constraint, i.e., modulo the second relation.
Once the strong constraint is solved so that the derivatives read as in
(\ref{strangeDs}) the second relation becomes
an identity\footnote{More explicitly, the left- and right-invariant Maurer-Cartan forms satisfy the identity
$e^{ia}\partial_ie_{a}{}^{j}-\bar{e}^{i\bar{a}}\partial_i\bar{e}_{\bar{a}}{}^{j}=0$. In order to verify this, one may use
(\ref{LRrel}) to reduce the left-hand side to $D^a\gamma_{a}{}^{\bar{a}}$, which vanishes as a consequence of  (\ref{gammaderrel}).} and hence the first relation holds without the need to invoke a constraint.
In ref.~\cite{Blumenhagen:2014gva} the results are presented in a different guise:
two independent partial derivatives $\partial_i$ and $\partial_{\bar{i}}$ are introduced,
corresponding to a doubled set of coordinates $(x^i, \bar{x}^{\bar{i}})$,
and the flattened derivatives are defined as
 \be\label{evenstrangerDs}
  D_{a} \ = \  e_{a}{}^{i}\partial_i\;, \qquad
  D_{\bar{a}} \ = \ \bar{e}_{\bar{a}}{}^{\bar{i}}\partial_{\bar{i}}\;.
 \ee
Provided, $e_{a}{}^{i}$ is assumed to depend only on $x$ and $\bar{e}_{\bar{a}}{}^{\bar{i}}$ to depend only on
$\bar{x}$,\footnote{Since for background group manifolds $e_{a}{}^{i}$ and
$\bar{e}_{\bar{a}}{}^{\bar{i}}$ depend on just one set of coordinates, the coordinates of the group manifold,
this doubling amounts to deviating from WZW backgrounds.
It is not clear to us in which sense one is then dealing with a
consistent string background CFT.}  the first of the relations (\ref{summREL}) holds identically for $F_{ab}{}^{c}$ and
$F_{\bar{a}\bar{b}}{}^{\bar{c}}$ given in (\ref{nonvanF}), without the need to invoke a constraint.
(Indeed, this can then be viewed as the anholonomy relation for a frame basis of the \textit{conventional} manifold $G\times G$;
we will return to this observation in the next section.)
However, in contrast to (\ref{strangeDs}),
the derivatives (\ref{evenstrangerDs}) do not satisfy the second relation in (\ref{summREL}),
which therefore has to be interpreted as a constraint on the derivatives, analogous to, but different from,
the strong constraint
in conventional DFT. A solution of this constraint is of course given by $x=\bar{x}$, in which case
it coincides with the solution $\tilde{\partial}^i=0$ of the usual strong constraint.
It is not clear to us whether the constraint, interpreted in this way, allows for other solutions.
We refrain from commenting further on this possibility but rather make some remarks on more
general proposals in the next section.

\section{Comments on non-geometry and background independence}
We now use the opportunity and ask the question whether the perturbative results
discussed above can teach us something about backgrounds that are genuinely non-geometric.
Such backgrounds can be motivated by considering gauged supergravity in lower dimensions,
whose gauge couplings are encoded in
structure constants or embedding tensors closely related to the background `fluxes'
$F_{ABC}$ and $F_A$ discussed above, subject to certain group-theoretical constraints.
It turns out these constraints allow for structure constants that cannot be obtained from a
generalized frame that is geometric in the sense of satisfying the strong constraint.
For instance, the constraint (\ref{Fsquareconst}), $F^{ABC}F_{ABC}=0$, following from
the strong constraint for $F_A=0$, need not be satisfied in gauged supergravity \cite{Dibitetto:2012rk}.
Turning on $\tilde{x}$ dependence, however, the more general fluxes can be obtained
in certain cases (although a complete classification seems to be lacking), but this requires
relaxing the strong constraint. Although it is now relatively well understood why the resulting
gauged supergravities obtained in this way, namely through generalized Scherk-Schwarz
compactification, are consistent despite violating the otherwise essential strong constraint \cite{Geissbuhler:2013uka},
we still do not have a proper understanding of the physical nature of these genuinely
extended spaces before compactification.

In the following we will make some general comments on this problem within the
perturbative framework, which allows one to sharpen
some of the issues.
Suppose we simply pick structure constants $F_{ABC}$ of the more general type by hand,
forgetting for the moment that they cannot be obtained from a generalized frame,
can we then write a consistent cubic theory (i.e.~one that is gauge invariant)?
The trouble with this idea is that the background not only enters the $F_{ABC}$ but
also the background flattened derivatives $D_A=\bar{E}_{A}{}^{M}\partial_M$,
and gauge invariance requires that they satisfy $[D_A,D_B]=F_{AB}{}^{C} D_{C}$.
This relation, with the $F_{ABC}$ defined in (\ref{FDef}), only holds modulo the strong constraint,
and thus relaxing the strong constraint is incompatible with gauge invariance.\footnote{It would
be interesting to investigate under which conditions one could achieve gauge invariance without
$[D_A,D_B]=F_{AB}{}^{C} D_{C}$ being satisfied in general. For instance,
in the context of Kaluza-Klein compactification one could take the background to depend on
doubled internal coordinates but the fluctuations only on external coordinates, in
which case the relation would hold on fluctuations, which for constant $F_{ABC}$
is sufficient for gauge invariance. As we are here interested in genuinely non-geometric theories
\textit{before} compactification, we will not discuss this possibility any further.}

The proposal put forward in~\cite{Blumenhagen:2015zma,Bosque:2015jda}
is to overcome this problem by changing the definition of the background frame field and
the associated background fluxes. Specifically, the background is assumed to be governed
by a \textit{conventional} frame field, which we denote here by ${\cal E}_{A}{}^{M}$,
and the $F_{AB}{}^{C}$ are defined to be the conventional coefficients of anholonomy,
 \be
  D_A \ = \ {\cal E}_{A}{}^{M}\partial_M\;, \qquad
  F_{AB}{}^{C} \ \equiv \ 2D_{[A}{\cal E}_{B]}{}^{M} {\cal E}_{CM}\;.
 \ee
The flattened derivatives thus defined then satisfy $[D_A,D_B]=F_{AB}{}^{C} D_{C}$
without the need to assume any constraints. In fact, in this form one is simply dealing with
standard geometry, but on a $2D$-dimensional space. In particular,
this definition of $F_{AB}{}^{C}$ is covariant under~\textit{conventional} diffeomorphisms on the
$2D$-dimensional manifold, not the generalized diffeomorphisms of DFT.
Consequently, in the proposal of \cite{Blumenhagen:2015zma,Bosque:2015jda}
there are two types of invariances: background transformations given by standard
diffeomorphisms and gauge transformations of the fluctuations governed by generalized
diffeomorphisms.

While it seems unnatural to have a formulation in which the background is described by
a conventional geometry but the fluctuations are described by a generalized geometry, the more serious issue
is whether such a theory can have the physical property of background independence.
More precisely, even for a theory for which one does not have a manifestly background independent
formulation one may ask whether the theory is background independent in the sense that any shift
of the background can be absorbed into a shift of the fluctuations, up to possible field
redefinitions.  For instance, although we do not yet have a manifestly background independent
formulation of string theory, it has been verified for bosonic string theory (for which a second-quantized
string field theory is available so that such questions can be meaningfully addressed)
that it is background independent w.r.t.~to certain marginal deformations \cite{Sen:1993mh}.

In order to make this point more transparent, let us discuss how the property of
background independence is realized in the original DFT written in the perturbative
form developed in this paper.
Here background independence is of course guaranteed in that
the split into background and fluctuations is completely arbitrary.
We may freely shift the background if we compensate this operation
by an opposite shift of the fluctuation.
More precisely, under shifts given by the variations
 \be\label{backgroundshifts}
 \begin{split}
  \delta_{\Delta} \bar{E}_{A}{}^{M} \ &= \ \Delta_{A}{}^{B}\bar{E}_{B}{}^{M}\;, \\[0.5ex]
  \delta_{\Delta} h_{A}{}^{B} \ &= \ \Delta_{A}{}^{B}-h_{A}{}^{C}\Delta_{C}{}^{B}\;,
 \end{split}
 \ee
the expansion ansatz
$E_{A}{}^{M} = \bar{E}_{A}{}^{M}-h_{A}{}^{B}\bar{E}_{B}{}^{M}$ is exactly invariant.
The background metric shifts as $\delta\bar{\cal G}_{AB}=2\Delta_{(AB)}$, so that the constraint
$\bar{\cal G}_{a\bar{b}}=0$ requires $\Delta_{a\bar{b}}=-\Delta_{\bar{b}a}$. Moreover,
the fluctuation $h_{AB}$ with one index lowered shifts as
 \be\label{backgroundshifts2}
  \delta_{\Delta} h_{AB} \ = \ \Delta_{AB}+h_{A}{}^{C} \Delta_{BC}\;.
 \ee
Since we determined the perturbation theory in terms of $h_{a\bar{b}}$ by
employing a gauge fixing condition, we have to take into account compensating
$GL(D)\times GL(D)$ transformations, in exactly the same way as for
the generalized diffeomorphisms. The combined action of the background shifts
(\ref{backgroundshifts2}) and the frame transformations on, say, $h_{ab}$ reads
 \be\label{compLOR}
  \delta h_{ab} \ = \ \Delta_{ab}+h_{a}{}^{\bar{c}}\Delta_{b\bar{c}}-\Lambda_{ab} \ = \ 0
  \quad \Rightarrow \quad \Lambda_{ab} \ = \ \Delta_{ab}+h_{a}{}^{\bar{c}} \Delta_{b\bar{c}}\;,
 \ee
which determines $\Lambda_{ab}$ so as to preserve the gauge condition $h_{ab}=0$.
A similar compensating transformation follows from $h_{\bar{a}\bar{b}}=0$.
We can now determine from (\ref{backgroundshifts2}) the full background shifts on the physical $h_{a\bar{b}}$,
 \be
  \delta h_{a\bar{b}} \ =  \ \Delta_{a\bar{b}}+h_{a}{}^{\bar{c}}\Delta_{\bar{b}\bar{c}}+\Lambda_{a}{}^{c} h_{c\bar{b}}\;,
 \ee
upon inserting the compensating frame transformation (\ref{compLOR}),
 \be\label{physicalShift}
  \delta_{\Delta} h_{a\bar{b}} \ = \ \Delta_{a\bar{b}}+h_{a}{}^{\bar{c}}\Delta_{\bar{b}\bar{c}}
  +\Delta_{a}{}^{c} h_{c\bar{b}} + h_{a\bar{c}}\, \Delta^{c\bar{c}}\, h_{c\bar{b}}\;.
 \ee
As for the gauge transformations, this is exact with no higher terms than quadratic in $h$. We thus conclude that arbitrary shifts of the background can be absorbed by the fluctuations, and viceversa, proving the full background independence of this formalism.

In order to further illustrate background independence, we now re-derive and extend from
these general variations the results obtained in \cite{Hohm:2010jy} for the class of flat toroidal
backgrounds. In this case the background frame field is most conveniently parameterized as
 \be\label{BackgroundFrameTorus}
  \bar{E}_{A}{}^{M} \ = \
   \begin{pmatrix}  \bar{E}_{ai} & \bar{E}_{a}{}^{i} \\[0.5ex]
   \bar{E}_{\bar{a}i} & \bar{E}_{\bar{a}}{}^{i} \end{pmatrix}
  \ = \  \begin{pmatrix}  -E_{ai}  & \delta_{a}{}^{i} \\[0.5ex]
   {E}_{i\bar{a}} & \delta_{\bar{a}}{}^{i} \end{pmatrix}\;,
 \ee
in terms of
 \be
  E_{ij} \  = \ G_{ij} + B_{ij}\;,
 \ee
where $G$ and $B$ are the constant background metric and B-field.
The tangent space metric then reads
 \be\label{tangentmetric}
  \bar{\cal G}_{AB} \ = \  \begin{pmatrix} -2\, G_{ab} & 0 \\[0.5ex]
  0 & 2\,G_{\bar{a}\bar{b}}\end{pmatrix}\;.
 \ee
In the above form we have fixed completely the \textit{background} $GL(D)\times GL(D)$
transformations by rotating $\bar{E}_{a}{}^{i}$ and $\bar{E}_{\bar{a}}{}^{i}$ into Kronecker deltas.
Note that this allows us to identify flat indices $a,b$ and $\bar{a},\bar{b}$ with curved indices
$i,j$, as used for the remaining entries $E_{ij}$ of the frame.
Due to the above background gauge fixing we
have to add further compensating gauge transformations, this time of the \textit{background}
$GL(D)\times GL(D)$ transformations with parameter $\bar{\Lambda}$.
Specifically, we need to preserve $\bar{E}_{a}{}^{i}=\delta_{a}{}^{i}$,
 \be
  \delta \bar{E}_{a}{}^{i} \ = \ \Delta_{a}{}^{b}\bar{E}_{b}{}^{i}+\Delta_{a}{}^{\bar{b}}\bar{E}_{\bar{b}}{}^{i}
  +\bar{\Lambda}_{a}{}^{b}\bar{E}_{b}{}^{i} \ = \ 0\;,
 \ee
and similarly for $\bar{E}_{\bar{a}}{}^{i}=\delta_{\bar{a}}{}^{i}$. This yields in total
 \be\label{COMPBACKframe}
  \begin{split}
   \bar{\Lambda}_{a}{}^{b} \ &= \ - \Delta_{a}{}^{b}-\Delta_{a}{}^{\bar{b}}\, \delta_{\bar{b}}{}^{b}\;, \\[0.5ex]
   \bar{\Lambda}_{\bar{a}}{}^{\bar{b}} \ &= \ -\Delta_{\bar{a}}{}^{\bar{b}} - \Delta_{\bar{a}}{}^{b}\, \delta_{b}{}^{\bar{b}}\;,
  \end{split}
 \ee
where we used a somewhat redundant notation with $\delta_{b}{}^{\bar{b}}$ and $\delta_{\bar{b}}{}^{b}$
in order to keep track of the index structure on $\Delta$.
We can then compute the background shifts of the non-trivial components of (\ref{BackgroundFrameTorus}),
 \be
 \begin{split}
  \delta \bar{E}_{ai} \ = \ -\delta  E_{ia} \  = \ -\Delta_{a}{}^{b} E_{bi} + \Delta_{a}{}^{\bar{b}} E_{i\bar{b}}
  -\bar{\Lambda}_{a}{}^{b} E_{bi}
  \ = \ \Delta_{a}{}^{\bar{b}}\big(E_{i\bar{b}}+E_{\bar{b}i}\big) \ = \ 2\Delta_{a}{}^{\bar{b}} G_{i\bar{b}}\;.
 \end{split}
 \ee
Recalling that the indices on $\Delta$ are raised and lowered with
(\ref{tangentmetric}) and setting $\chi_{a\bar{b}}\equiv \Delta_{a\bar{b}}$ we obtain
 \be\label{Eshift}
  \delta E_{ij} \ =  \ -\chi_{ij}\;.
 \ee
The analogous computation for $\bar{E}_{\bar{a}i}$ yields the same result.
Similarly, we can compute the shift of $h_{a\bar{b}}$ by adding to (\ref{physicalShift})
background frame transformations (\ref{BACkgroundFRame}) with parameters (\ref{COMPBACKframe}),
 \be
  \begin{split}
   \delta_{\Delta} h_{a\bar{b}} \ &= \ \Delta_{a\bar{b}}+h_{a}{}^{\bar{c}}\Delta_{\bar{b}\bar{c}}
  +\Delta_{a}{}^{c} h_{c\bar{b}} + h_{a\bar{c}}\, \Delta^{c\bar{c}}\, h_{c\bar{b}}
  +\bar{\Lambda}_{a}{}^{c} h_{c\bar{b}} +\bar{\Lambda}_{\bar{b}}{}^{\bar{c}} h_{a\bar{c}} \\[0.5ex]
  \ &= \ \Delta_{a\bar{b}}-\Delta_{a}{}^{\bar{c}}\,\delta_{\bar{c}}{}^{c}\, h_{c\bar{b}}
  +\Delta^{c}{}_{\bar{b}} \, \delta_{c}{}^{\bar{c}} \,h_{a\bar{c}}
  + h_{a\bar{c}}\, \Delta^{c\bar{c}}\, h_{c\bar{b}}\;,
 \end{split}
 \ee
where we used $\Delta_{\bar{b}c}=-\Delta_{c\bar{b}}$.
In order to compare this with the analysis of background independence in \cite{Hohm:2010jy}
we use the result from \cite{Hohm:2011dz} that the string field theory variable $e_{ij}$ around flat space
with generalized frame (\ref{BackgroundFrameTorus})
coincides with $h_{a\bar{b}}$ to all orders, upon identifying indices by means of the trivial
vielbeins $\delta_{i}{}^{a}$ and $\delta_{i}{}^{\bar{a}}$. Converting then the above equation into curved indices
we read off
 \be\label{finaldeltae}
  \delta e_{ij} \ = \ \chi_{ij}-\tfrac{1}{2}\,\chi_{i}{}^{k} \, e_{kj} -\tfrac{1}{2}\, \chi^{k}{}_{j}\,e_{ik}
  -\tfrac{1}{4}\, e_{il}\, \chi^{kl}\, e_{kj}\;,
 \ee
where we took into account the relative factors of $\pm \tfrac{1}{2}$ originating from (\ref{tangentmetric})
when converting contractions with $\bar{\cal G}$ to contractions with $G$. Together with (\ref{Eshift})
and to linear order in fields this agrees precisely with the result implied by eq.~(2.7)
in \cite{Hohm:2010jy}.\footnote{This corrects a typo in eqs.~(2.8) and (2.15) of \cite{Hohm:2010jy}.}
The quadratic or higher contributions were not determined in \cite{Hohm:2010jy}, but here we obtained
the exact result, which shows that there are no terms higher than quadratic in fields needed.\footnote{As a
consistency check one may verify that the full background independent variable \cite{Hohm:2010jy},
 \be
  {\cal E} \ \equiv \  E + \big(1-\tfrac{1}{2}eG^{-1}\big)^{-1}e
 \ee
satisfies $\delta {\cal E}=0$ under (\ref{Eshift}) and (\ref{finaldeltae}).}

Let us emphasize that in the above analysis of background independence we have to take the background shifts
$\chi_{ij}$ to be constant in order to stay within the class of constant backgrounds (\ref{BackgroundFrameTorus}).
This is necessary because the original cubic action of
\cite{Hull:2009mi} is only valid for flat backgrounds. Of course, in the full
background independent DFT we can perform arbitrary shifts, but this would switch on, for instance, connection
terms that have been set to zero for flat backgrounds. Is there a similar property of (restricted)
background independence for the cubic action on WZW backgrounds?
This case is more restrictive because the employed metric and 3-form $H$ are uniquely determined once
the Lie algebra $f_{ab}{}^{c}$ has been fixed. Thus, in contrast to toroidal backgrounds on which one can
put arbitrary constant metrics and $B$-fields, for the WZW backgrounds there is no simple property of
background independence visible for the cubic action. Again, in the full background independent
DFT we can perform arbitrary shifts away from the WZW backgrounds, but this would generally switch
on curvature and connection terms (for instance $\omega_{a\bar{b}\bar{c}}$) that were set to zero before.

After this digression into the background independence property of particular backgrounds,
let us return to the general discussion. A background frame $\bar{E}_{A}{}^{M}$ a priori
carries $(2D)^2=4D^2$ components, but in order to determine the number of
physical (i.e.~gauge inequivalent) background shifts we have to take into account
the (background) local frame transformations. These are given by $GL(D)\times GL(D)$,
which eliminates $D^2+D^2$ degrees of freedom. In addition, we have the constraint
$\bar{\cal G}_{a\bar{b}}=\bar{E}_{a}{}^{M}\bar{E}_{\bar{b}M}=0$, which eliminates another
$D^2$ components, leaving $D^2$ physical degrees of freedom, which can be shifted arbitrarily
by the $\Delta_{a\bar{b}}$. This matches precisely the components of  $h_{a\bar{b}}$,
as needed for background independence.
In contrast, taking the background to be given by a conventional frame field ${\cal E}_{A}{}^{M}$
there is no constraint corresponding to $\bar{\cal G}_{a\bar{b}}=0$ and hence
we have more background shifts than can be absorbed into a shift of the physical fluctuation
$h_{a\bar{b}}$. Could we pose an additional constraint on ${\cal E}_{A}{}^{M}$
so that background and fluctuations carry the same number of components?
This is not possible because the metric $\eta_{MN}$ was needed
in order to define the constraint, but in standard geometry there is no diffeomorphism invariant metric
available. The reason that the constraint $\eta_{MN}\bar{E}_{a}{}^{M}\bar{E}_{\bar{b}}{}^{N}=0$ can be consistently
imposed in DFT is precisely that the geometry is governed by \textit{generalized} diffeomorphisms,
under which $\eta_{MN}$ \textit{is} invariant.
Summarizing, background independence
for a general class of backgrounds
can be realized if \textit{both}
the background and the fluctuations are governed by the same geometries but
is problematic if the background and fluctuations are governed by different geometries.

\section*{Acknowledgments}
We would like to thank the organizers of the CERN workshop ``Duality Symmetries in String and
M-theories", where this project was initiated.
We have benefited from comments by Ralph Blumenhagen, Pascal du Bosque, Falk Hassler
and Dieter L\"ust explaining their work. We would also like to thank  Warren Siegel and Barton Zwiebach
for useful discussions and Ergin Sezgin for encouraging us to perform the
general background expansion in DFT.
 The work of O.H. is supported by a DFG Heisenberg fellowship. The work of D.M. is supported by CONICET.


\section*{Appendix}

\begin{appendix}

\section{Further relations for cubic action on general backgrounds}

In this appendix we collect a few explicit expressions for the cubic action that may be more
convenient for applications. Moreover, below we derive equivalent expressions using the
strong constraint, which have the advantage that unbarred and barred indices are treated on the
same footing.
First, separating the action into quadratic and cubic parts,
\be
S = \int dX e^{-2 \bar d} \left({\cal L}_2 + {\cal L}_3\right)\;,  \label{CubicActionFluxes}
\ee
we determine from (\ref{finalcubicstuff}) for the quadratic part
\begin{eqnarray}\label{explicitA1}
{\cal L}_2 \!\!&=&\!\!  - 8\, {D}^{a}{{D}_{a}{d}\, }\,  d - 8\, {D}^{a}{{D}^{\bar a}{{h}_{a \bar a}}\, }\,  d - 2\, {D}^{a}{{D}_{a}{{h}^{b \bar a}}\, }\,  {h}_{b \bar a} + 4\, {D}^{a}{{D}^{b}{{h}_{a}\,^{\bar a}}\, }\,  {h}_{b \bar a}  - 4\, {D}^{a}{{D}^{b}{{h}_{b}\,^{\bar a}}\, }\,  {h}_{a \bar a}  \nonumber \\ &&
- 2\, {D}^{a}{{h}_{a}\,^{\bar a}}\,  {D}^{b}{{h}_{b \bar a}}\,  + 2\, {D}^{\bar a}{{h}^{a}\,_{\bar a}}\,  {D}^{\bar b}{{h}_{a \bar b}}\,
 \nonumber \\[1ex] &&
- 8\, {D}^{a}{d}\,  {F}_{a} d
  - 8\, {D}^{a}{{h}_{a}\,^{\bar a}}\,  {F}_{\bar a} d   - 8\, {D}^{\bar a}{{h}^{a}\,_{\bar a}}\,  {F}_{a} d  - 4\, {D}^{a}{{h}_{a}\,^{\bar a}}\,  {F}^{b} {h}_{b \bar a} - 2\, {D}^{a}{{h}^{b \bar a}}\,  {F}_{a} {h}_{b \bar a} \nonumber \\ &&+ 8\, {D}^{a}{{h}^{b \bar a}}\,  {F}_{a b \bar a} d + 4\, {D}^{\bar a}{{h}^{a}\,_{\bar a}}\,  {F}^{\bar b} {h}_{a \bar b}  - 4\, {D}^{a}{{h}^{b \bar a}}\,  {F}_{a b}\,^{c} {h}_{c \bar a} - 4\, {D}^{a}{{h}_{a}\,^{\bar a}}\,  {F}_{\bar a}\,^{b \bar b} {h}_{b \bar b} \nonumber \\ &&  - 4\, {D}^{a}{{h}^{b \bar a}}\,  {F}_{a \bar a}\,^{\bar b} {h}_{b \bar b} - 4\, {D}^{\bar a}{{h}^{a}\,_{\bar a}}\,  {F}_{a}\,^{b \bar b} {h}_{b \bar b}
    \nonumber \\[1ex] &&
     - 8\, {D}^{a}{{F}^{\bar a}}\,  d \, {h}_{a \bar a}+ 8\, {D}^{a}{{F}_{a}\,^{b \bar a}}\,  d\,  {h}_{b \bar a}
 + 8\, {D}^{a}{{F}^{b \bar a \bar b}}\,  {h}_{a \bar a} {h}_{b \bar b}
  \nonumber \\[0.5ex] &&
    - 8\, {F}^{a} {F}^{\bar a} d {h}_{a \bar a}
  - 2\, {F}^{a} {F}^{b} {h}_{a}\,^{\bar a} {h}_{b \bar a} + 8\, {F}^{a} {F}_{a}\,^{b \bar a} d {h}_{b \bar a} + 2\, {F}^{\bar a} {F}^{\bar b} {h}^{a}\,_{\bar a} {h}_{a \bar b} \nonumber \\ && + 4\, {F}^{a} {F}^{b \bar a \bar b} {h}_{a \bar a} {h}_{b \bar b} - 4\, {F}^{\bar a} {F}^{a b \bar b} {h}_{a \bar a} {h}_{b \bar b} - 4\, {F}^{a b c} {F}_{a}\,^{\bar a \bar b} {h}_{b \bar a} {h}_{c \bar b} + 2\, {F}^{a b \bar a} {F}_{a \bar a}\,^{c} {h}_{b}\,^{\bar b} {h}_{c \bar b} \nonumber \\ &&  + 2\, {F}^{a b \bar a} {F}_{a}\,^{c \bar b} {h}_{b \bar a} {h}_{c \bar b} + 2\, {F}^{a \bar a \bar b} {F}_{a \bar a}\,^{\bar c} {h}^{b}\,_{\bar b} {h}_{b \bar c} - 2\, {F}^{a \bar a \bar b} {F}_{\bar a}\,^{b \bar c} {h}_{a \bar b} {h}_{b \bar c} + 4\, {F}^{a \bar a \bar b} {F}_{\bar a}\,^{b \bar c} {h}_{a \bar c} {h}_{b \bar b} \; , \ \ \ \ \ \ \ \
\end{eqnarray}
and for the cubic part
\begin{eqnarray}\label{explicitA2}
{\cal L}_3 \!\!&=&\!\! 8\, {D}^{a}{{D}_{a}{d}\, }\,  {d}^{2} + 8\, {D}^{a}{d}\,  {F}_{a} {d}^{2} + 16\, {D}^{\bar a}{{D}^{a}{d}\, }\,  d {h}_{a \bar a} - 8\, {D}^{a}{{D}^{b}{{h}_{a}\,^{\bar a}}\, }\,  d {h}_{b \bar a} - 4\, {D}^{a}{{h}_{a}\,^{\bar a}}\,  {D}^{b}{{h}_{b \bar a}}\,  d \nonumber \\ && - 4\, {D}^{a}{{h}^{b \bar a}}\,  {D}_{a}{{h}_{b \bar a}}\,  d + 8\, {D}^{\bar a}{{D}^{\bar b}{{h}^{a}\,_{\bar b}}\, }\,  d {h}_{a \bar a} + 4\, {D}^{\bar a}{{h}^{a}\,_{\bar a}}\,  {D}^{\bar b}{{h}_{a \bar b}}\,  d + 4\, {D}^{a}{{D}^{\bar a}{{h}^{b \bar b}}\, }\,  {h}_{a \bar b} {h}_{b \bar a} \nonumber \\ &&- 8\, {D}^{a}{{F}^{b}}\,  d {h}_{a}\,^{\bar a} {h}_{b \bar a} + 16\, {D}^{a}{d}\,  {F}_{a}\,^{b \bar a} d {h}_{b \bar a} + 4\, {D}^{a}{{h}^{b \bar a}}\,  {D}^{\bar b}{{h}_{a \bar a}}\,  {h}_{b \bar b} + 4\, {D}^{a}{{h}^{b \bar a}}\,  {D}_{\bar a}{{h}_{b}\,^{\bar b}}\,  {h}_{a \bar b}\nonumber \\ && - 4\, {D}^{a}{{h}^{b \bar a}}\,  {D}^{\bar b}{{h}_{b \bar a}}\,  {h}_{a \bar b} - 8\, {D}^{a}{{h}_{a}\,^{\bar a}}\,  {F}^{b} d {h}_{b \bar a} - 8\, {D}^{a}{{h}^{b \bar a}}\,  {F}_{b} d {h}_{a \bar a} - 4\, {D}^{\bar a}{{D}^{a}{{h}^{b \bar b}}\, }\,  {h}_{a \bar b} {h}_{b \bar a} \nonumber \\ && + 8\, {D}^{\bar a}{{F}^{\bar b}}\,  d {h}^{a}\,_{\bar a} {h}_{a \bar b} + 8\, {D}^{\bar a}{{h}^{a}\,_{\bar a}}\,  {F}^{\bar b} d {h}_{a \bar b} + 8\, {D}^{\bar a}{{h}^{a \bar b}}\,  {F}_{\bar b} d {h}_{a \bar a} - 2\, {D}^{a}{{F}^{\bar a}}\,  {h}_{a}\,^{\bar b} {h}^{b}\,_{\bar a} {h}_{b \bar b} \nonumber \\ &&  - 8\, {D}^{a}{{F}^{b \bar a \bar b}}\,  d {h}_{a \bar a} {h}_{b \bar b} + 8\, {D}^{a}{{h}^{b \bar a}}\,  {F}_{a b}\,^{c} d {h}_{c \bar a} + 8\, {D}^{a}{{h}^{b \bar a}}\,  {F}_{a \bar a}\,^{\bar b} d {h}_{b \bar b} - 8\, {D}^{a}{{h}^{b \bar a}}\,  {F}_{b \bar a}\,^{\bar b} d {h}_{a \bar b} \nonumber \\ && - 8\, {D}^{\bar a}{{F}^{a b \bar b}}\,  d {h}_{a \bar a} {h}_{b \bar b} - 8\, {D}^{\bar a}{{h}^{a \bar b}}\,  {F}_{a \bar b}\,^{b} d {h}_{b \bar a} - 4\, {F}^{a} {F}^{b} d {h}_{a}\,^{\bar a} {h}_{b \bar a} + 4\, {F}^{\bar a} {F}^{\bar b} d {h}^{a}\,_{\bar a} {h}_{a \bar b} \nonumber \\ && - 8\, {D}^{a}{{h}^{b \bar a}}\,  {F}_{a}\,^{c \bar b} {h}_{b \bar b} {h}_{c \bar a} + 4\, {D}^{a}{{h}^{b \bar a}}\,  {F}_{b \bar a}\,^{c} {h}_{a}\,^{\bar b} {h}_{c \bar b} + 4\, {D}^{a}{{h}^{b \bar a}}\,  {F}_{b}\,^{c \bar b} {h}_{a \bar b} {h}_{c \bar a} - 4\, {D}^{a}{{h}^{b \bar a}}\,  {F}_{\bar a}\,^{\bar b \bar c} {h}_{a \bar b} {h}_{b \bar c}\nonumber \\ &&  + 2\, {D}^{\bar a}{{F}_{\bar a}\,^{a \bar b}}\,  {h}_{a}\,^{\bar c} {h}^{b}\,_{\bar b} {h}_{b \bar c} - 4\, {D}^{\bar a}{{h}^{a \bar b}}\,  {F}_{a}\,^{b c} {h}_{b \bar a} {h}_{c \bar b} - 4\, {D}^{\bar a}{{h}^{a \bar b}}\,  {F}_{a \bar b}\,^{\bar c} {h}^{b}\,_{\bar a} {h}_{b \bar c} - 4\, {D}^{\bar a}{{h}^{a \bar b}}\,  {F}_{\bar b}\,^{b \bar c} {h}_{a \bar c} {h}_{b \bar a} \nonumber \\ && + 2\, {F}^{\bar a} {F}_{\bar a}\,^{a \bar b} {h}_{a}\,^{\bar c} {h}^{b}\,_{\bar b} {h}_{b \bar c} + 8\, {F}^{a b c} {F}_{a}\,^{\bar a \bar b} d {h}_{b \bar a} {h}_{c \bar b} - 4\, {F}^{a b \bar a} {F}_{a \bar a}\,^{c} d {h}_{b}\,^{\bar b} {h}_{c \bar b} - 4\, {F}^{a b \bar a} {F}_{a}\,^{c \bar b} d {h}_{b \bar a} {h}_{c \bar b} \nonumber \\ && - 4\, {F}^{a \bar a \bar b} {F}_{a \bar a}\,^{\bar c} d {h}^{b}\,_{\bar b} {h}_{b \bar c} + 4\, {F}^{a \bar a \bar b} {F}_{\bar a}\,^{b \bar c} d {h}_{a \bar b} {h}_{b \bar c} - 8\, {F}^{a \bar a \bar b} {F}_{\bar a}\,^{b \bar c} d {h}_{a \bar c} {h}_{b \bar b} + 4\, {F}^{a b c} {F}_{a}\,^{d \bar a} {h}_{b \bar a} {h}_{c}\,^{\bar b} {h}_{d \bar b} \nonumber \\ && + \frac{4}{3}\, {F}^{a b c} {F}^{\bar a \bar b \bar c} {h}_{a \bar a} {h}_{b \bar b} {h}_{c \bar c} - 2\, {F}^{a b \bar a} {F}_{a \bar a}\,^{\bar b} {h}_{b}\,^{\bar c} {h}^{c}\,_{\bar b} {h}_{c \bar c} + 4\, {F}^{a b \bar a} {F}_{a}\,^{\bar b \bar c} {h}_{b \bar b} {h}^{c}\,_{\bar a} {h}_{c \bar c} \nonumber \\ &&- 4\, {F}^{a b \bar a} {F}_{\bar a}\,^{c \bar b} {h}_{a \bar b} {h}_{b}\,^{\bar c} {h}_{c \bar c} + 4\, {F}^{a b \bar a} {F}^{c \bar b \bar c} {h}_{a \bar b} {h}_{b \bar c} {h}_{c \bar a} + 4\, {F}^{a \bar a \bar b} {F}_{\bar a}\,^{\bar c \bar d} {h}_{a \bar c} {h}^{b}\,_{\bar b} {h}_{b \bar d} \; .
\end{eqnarray}

\bigskip

Next,  we would like to express the covariant background expanded generalized Ricci scalar and action in a barred-unbarred democratic form. Note that we have started from the background independent frame-like generalized Ricci scalar (\ref{FullGenRicci}) that is not democratic in barred and unbarred flat indices. However, recalling that the following combination vanishes due to the strong constraint (see for example page 17 of \cite{Siegel:1993th})
\begin{eqnarray}
{\cal Z} &=& -4 \left( E_A  {\cal F}^A + \frac 1 2 {\cal F}_A^2 + \frac 1 2 E_A E_B {\cal G}^{A B} - \frac 1 {12}  \Omega_{[A B C]}^2 + \frac 1 8 E^A {\cal G}^{B C} E_{B}{\cal G}_{A C}\right)\ =\ 0 \ ,
 \end{eqnarray}
one can bring the original action to a fully democratic form as follows
\begin{eqnarray}
{\cal R} = {\cal R} - \frac 1 2 {\cal Z} &=& -2 \left(E_a  {\cal F}^a - E_{\bar a} {\cal F}^{\bar a}\right) - \left(   {\cal F}_a^2 -   {\cal F}_{\bar a}^2\right) - \left(E_a E_b {\cal G}^{a b} - E_{\bar a} E_{\bar b} {\cal G}^{\bar a \bar b}\right) \\ &&- \frac 1 {4} \left(E^a {\cal G}^{b c} E_b {\cal G}_{a c} - E^{\bar a} {\cal G}^{\bar b \bar c} E_{\bar b} {\cal G}_{\bar a \bar c}\right)
+ \frac 1 2  \left(   \Omega_{ab \bar c}^2 -   \Omega_{\bar a \bar b c}^2\right) + \frac 1 {6} \left(  \Omega_{[abc]}^2 -   \Omega_{[\bar a \bar b \bar c]}^2 \right)\nonumber \ ,
\end{eqnarray}
where we used the following identities
\begin{eqnarray}
 \Omega_{ab\bar c} &=&  \Omega_{[a b \bar c]} \ ,\\
 \Omega_{\bar a \bar b  c} &=&  \Omega_{[\bar a \bar b c]} \ ,\\
 \Omega_{[ABC]}^2 &=&  \Omega_{[abc]}^2 +  \Omega_{[\bar a \bar b \bar c]}^2 + 3  \Omega_{[a b \bar c]}^2 + 3 \Omega_{[\bar a \bar b c]}^2 \ .
\end{eqnarray}
Then, by background expanding $\cal Z$ we can democratize order by order the generalized Ricci scalar. The background covariant expansion of $\cal Z$ (which, as we mentioned above, vanishes due to the strong constraint) is given by
\begin{eqnarray}
{\cal Z}_0 &=& - 2\, {\bar {\cal R}}^{a b}\,_{b a} - 2\, {\bar {\cal R}}^{\bar a \bar b}\,_{\bar b \bar a} \ , \\
{\cal Z}_1 &=& 8\, {\bar \nabla}^{a}{{\bar \nabla}_{a}{d}\, }\,  + 8\, {\bar \nabla}^{\bar a}{{\bar \nabla}_{\bar a}{d}\, }\,  + 4\, {\bar \nabla}^{a}{{\bar \nabla}^{\bar a}{{h}_{a \bar a}}\, }\,  - 4\, {\bar \nabla}^{\bar a}{{\bar \nabla}^{a}{{h}_{a \bar a}}\, }\,  + 8\, {\bar {\cal R}}^{a b \bar a}\,_{a} {h}_{b \bar a} - 8\, {\bar {\cal R}}^{\bar a \bar b a}\,_{\bar a} {h}_{a \bar b} \ , \\
{\cal Z}_2 &=&  - 8\, {\bar \nabla}^{a}{d}\,  {\bar \nabla}_{a}{d}\,  - 8\, {\bar \nabla}^{\bar a}{d}\,  {\bar \nabla}_{\bar a}{d}\,  + 2\, {\bar \nabla}^{a}{{h}^{b \bar a}}\,  {\bar \nabla}_{a}{{h}_{b \bar a}}\, + 2\, {\bar \nabla}^{\bar a}{{h}^{a \bar b}}\,  {\bar \nabla}_{\bar a}{{h}_{a \bar b}}\,  \nonumber \\ && - 4\, {h}_{a \bar a} {\bar \nabla}^{a}{{\bar \nabla}^{b}{{h}_{b}\,^{\bar a}}\, }\,  + 4\, {h}_{a \bar a} {\bar \nabla}^{b}{{\bar \nabla}^{a}{{h}_{b}\,^{\bar a}}\, }\,  - 4\, {h}_{a \bar a} {\bar \nabla}^{\bar a}{{\bar \nabla}^{\bar b}{{h}^{a}\,_{\bar b}}\, }\,  + 4\, {h}_{a \bar a} {\bar \nabla}^{\bar b}{{\bar \nabla}^{\bar a}{{h}^{a}\,_{\bar b}}\, }\, \nonumber \\ && + 4\, {h}_{a}\,^{\bar a} {h}_{b \bar a} {\bar {\cal R}}^{c a b}\,_{c} + 8\, {h}_{a \bar a} {h}_{b \bar b} {\bar {\cal R}}^{a b \bar a \bar b} + 4\, {h}^{a}\,_{\bar a} {h}_{a \bar b} {\bar {\cal R}}^{\bar c \bar a \bar b}\,_{\bar c} \ , \\ \nonumber
{\cal Z}_3 &=&  - 2\, {h}_{a}\,^{\bar a} {h}_{b \bar a} {\bar \nabla}^{a}{{\bar \nabla}^{\bar b}{{h}^{b}\,_{\bar b}}\, }\,  - 2\, {h}^{a}\,_{\bar a} {h}_{a \bar b} {\bar \nabla}^{b}{{\bar \nabla}^{\bar a}{{h}_{b}\,^{\bar b}}\, }\,  + 2\, {h}_{a}\,^{\bar a} {h}_{b \bar a} {\bar \nabla}^{\bar b}{{\bar \nabla}^{a}{{h}^{b}\,_{\bar b}}\, }\,  + 2\, {h}^{a}\,_{\bar a} {h}_{a \bar b} {\bar \nabla}^{\bar a}{{\bar \nabla}^{b}{{h}_{b}\,^{\bar b}}\, }\, \nonumber\\ && - 4\, {h}_{a}\,^{\bar a} {h}_{b \bar b} {h}_{c \bar a} {\bar {\cal R}}^{a b \bar b c} - 4\, {h}_{a}\,^{\bar a} {h}^{b}\,_{\bar b} {h}_{b \bar a} {\bar {\cal R}}^{c a \bar b}\,_{c} + 4\, {h}_{a}\,^{\bar a} {h}^{b}\,_{\bar b} {h}_{b \bar a} {\bar {\cal R}}^{\bar c \bar b a}\,_{\bar c} + 4\, {h}_{a \bar a} {h}^{b}\,_{\bar b} {h}_{b \bar c} {\bar {\cal R}}^{\bar b \bar a a \bar c} \, . \ \ \ \ \ \ \ \ \
\end{eqnarray}

With this we can now write the background expansion of the generalized Ricci scalar in a democratic form
\begin{eqnarray}
{\cal R}_0 &=& - {\bar {\cal R}}^{a b}\,_{b a} + {\bar {\cal R}}^{\bar a \bar b}\,_{\bar b \bar a}\;, \\
{\cal R}_1 &=& 4\, {\bar \nabla}^{a}{{\bar \nabla}_{a}{d}\, }\, - 4\, {\bar \nabla}^{\bar a}{{\bar \nabla}_{\bar a}{d}\, }\, + 2\, {\bar \nabla}^{a}{{\bar \nabla}^{\bar a}{{h}_{a \bar a}}\, }\,  + 2\, {\bar \nabla}^{\bar a}{{\bar \nabla}^{a}{{h}_{a \bar a}}\, }\, \\
&& + 4\, {\bar {\cal R}}^{a b \bar a}\,_{a} {h}_{b \bar a} + 4\, {\bar {\cal R}}^{\bar a \bar b a}\,_{\bar a} {h}_{a \bar b}\;, \nonumber\\
{\cal R}_2 &=&  - 4\, {\bar \nabla}^{a}{d}\,  {\bar \nabla}_{a}{d}\, + 4\, {\bar \nabla}^{\bar a}{d}\,  {\bar \nabla}_{\bar a}{d}\, - 8\, {\bar \nabla}^{a}{d}\,  {\bar \nabla}^{\bar a}{{h}_{a \bar a}}\, - 8\, {\bar \nabla}^{\bar a}{d}\,{\bar \nabla}^{a}{{h}_{a \bar a}}\, \\&&
 + 2\, {\bar \nabla}^{a}{{\bar \nabla}^{b}{{h}_{a}\,^{\bar a}}\, }\,  {h}_{b \bar a} - 2\, {\bar \nabla}^{\bar a}{{\bar \nabla}^{\bar b}{{h}^{a}\,_{\bar a}}\, }\,  {h}_{a \bar b}+ 2\, {\bar \nabla}^{a}{{\bar \nabla}^{b}{{h}_{b}\,^{\bar a}}\, }\,  {h}_{a \bar a}  - 2\, {\bar \nabla}^{\bar a}{{\bar \nabla}^{\bar b}{{h}^{a}\,_{\bar b}}\, }\,  {h}_{a \bar a} \nonumber \\ && - 8\, {\bar \nabla}^{a}{{\bar \nabla}^{\bar a}{d}\, }\,  {h}_{a \bar a} - 8\, {\bar \nabla}^{\bar a}{{\bar \nabla}^{a}{d}\, }\,  {h}_{a \bar a} + 2\, {\bar \nabla}^{a}{{h}_{a}\,^{\bar a}}\,  {\bar \nabla}^{b}{{h}_{b \bar a}}\, - 2\, {\bar \nabla}^{\bar a}{{h}^{a}\,_{\bar a}}\,  {\bar \nabla}^{\bar b}{{h}_{a \bar b}}\, \nonumber \\&&  - {\bar \nabla}^{\bar a}{{h}^{a \bar b}}\,  {\bar \nabla}_{\bar a}{{h}_{a \bar b}}\, + {\bar \nabla}^{a}{{h}^{b \bar a}}\,  {\bar \nabla}_{a}{{h}_{b \bar a}}\, + 2\, {\bar {\cal R}}^{a b c}\,_{a} {h}_{b}\,^{\bar a} {h}_{c \bar a} - 2\, {\bar {\cal R}}^{\bar a \bar b \bar c}\,_{\bar a} {h}^{a}\,_{\bar b} {h}_{a \bar c} \;, \nonumber  \\
{\cal R}_3 &=&  - 8\, {\bar \nabla}^{a}{d}\,  {\bar \nabla}^{b}{{h}_{a}\,^{\bar a}}\,  {h}_{b \bar a} + 8\, {\bar \nabla}^{\bar a}{d}\,  {\bar \nabla}^{\bar b}{{h}^{a}\,_{\bar a}}\,  {h}_{a \bar b}- 8\, {\bar \nabla}^{a}{d}\,  {\bar \nabla}^{b}{{h}_{b}\,^{\bar a}}\,  {h}_{a \bar a} + 8\, {\bar \nabla}^{\bar a}{d}\,  {\bar \nabla}^{\bar b}{{h}^{a}\,_{\bar b}}\,  {h}_{a \bar a} \\&&+ 16\, {\bar \nabla}^{a}{d}\,  {\bar \nabla}^{\bar a}{d}\,  {h}_{a \bar a} - 8\, {\bar \nabla}^{a}{{\bar \nabla}^{b}{d}\, }\,  {h}_{a}\,^{\bar a} {h}_{b \bar a} + 8\, {\bar \nabla}^{\bar a}{{\bar \nabla}^{\bar b}{d}\, }\,  {h}^{a}\,_{\bar a} {h}_{a \bar b} \nonumber \\&& - {\bar \nabla}^{a}{{\bar \nabla}^{\bar a}{{h}_{a}\,^{\bar b}}\, }\,  {h}^{b}\,_{\bar a} {h}_{b \bar b} - {\bar \nabla}^{\bar a}{{\bar \nabla}^{a}{{h}^{b}\,_{\bar a}}\, }\,  {h}_{a}\,^{\bar b} {h}_{b \bar b} - 3\, {\bar \nabla}^{a}{{\bar \nabla}^{\bar a}{{h}^{b}\,_{\bar a}}\, }\,  {h}_{a}\,^{\bar b} {h}_{b \bar b} - 3\, {\bar \nabla}^{\bar a}{{\bar \nabla}^{a}{{h}_{a}\,^{\bar b}}\, }\,  {h}^{b}\,_{\bar a} {h}_{b \bar b}\nonumber \\ && - 2\, {\bar \nabla}^{a}{{\bar \nabla}^{\bar a}{{h}^{b \bar b}}\, }\,  {h}_{a \bar b} {h}_{b \bar a} - 2\, {\bar \nabla}^{\bar a}{{\bar \nabla}^{a}{{h}^{b \bar b}}\, }\,  {h}_{a \bar b} {h}_{b \bar a} - 4\, {\bar \nabla}^{a}{{h}_{a}\,^{\bar a}}\,  {\bar \nabla}^{\bar b}{{h}^{b}\,_{\bar b}}\,  {h}_{b \bar a} \nonumber \\ && - 4\, {\bar \nabla}^{a}{{h}_{a}\,^{\bar a}}\,  {\bar \nabla}^{\bar b}{{h}^{b}\,_{\bar a}}\,  {h}_{b \bar b}\, - 4\, {\bar \nabla}^{a}{{h}^{b \bar a}}\,  {\bar \nabla}^{\bar b}{{h}_{b \bar b}}\,  {h}_{a \bar a} \, - 4\, {\bar \nabla}^{a}{{h}^{b \bar a}}\,  {\bar \nabla}_{\bar a}{{h}_{a}\,^{\bar b}}\,  {h}_{b \bar b} - 4\, {\bar \nabla}^{a}{{h}^{b \bar a}}\,  {\bar \nabla}^{\bar b}{{h}_{b \bar a}}\,  {h}_{a \bar b} \nonumber \\ && - 2\, {\bar {\cal R}}^{a b \bar a}\,_{a} {h}_{b}\,^{\bar b} {h}^{c}\,_{\bar a} {h}_{c \bar b}
 - 2\, {\bar {\cal R}}^{\bar a \bar b a}\,_{\bar a} {h}_{a}\,^{\bar c} {h}^{b}\,_{\bar b} {h}_{b \bar c} - 2\, {\bar {\cal R}}^{a b \bar a c} {h}_{a}\,^{\bar b} {h}_{b \bar a} {h}_{c \bar b}- 2\, {\bar {\cal R}}^{\bar a \bar b a \bar c} {h}_{a \bar b} {h}^{b}\,_{\bar a} {h}_{b \bar c} \ . \nonumber
\end{eqnarray}

With this information, we can then follow the procedure of the previous section in order to compute the democratic form of the cubic action, after simplifying it with the assumption that  the background satisfies its equations of motion, and performing integration by parts. The final result is given by
\begin{eqnarray}
S = \int dX\, e^{- 2 \bar d}\, \left( {\cal L}_2 + {\cal L}_3  \right) \, , \nonumber
\end{eqnarray}
with
\begin{eqnarray}
{\cal L}_2 &=& - \bar \nabla^a \bar \nabla_a h^{b \bar a} h_{b \bar a} + \bar \nabla^{\bar a} \bar \nabla_{\bar a} h^{a \bar b} h_{a \bar b} - 2 \bar \nabla^a h_{a \bar a} \bar \nabla_b h^{b \bar a} + 2 \bar \nabla^{\bar a} h_{a \bar a} \bar \nabla_{\bar b} h^{a \bar b} \nonumber\\ && - 4 \bar \nabla^{[b} \bar \nabla^{a]}h_a{}^{\bar a} h_{b \bar a} + 4 \bar \nabla^{[\bar b} \bar \nabla^{\bar a]}h^a{}_{\bar a} h_{a \bar b} + 2 \bar {\cal R}^{a b c}{}_a h_b{}^{\bar a} h_{c \bar a} - 2 \bar {\cal R}^{\bar a \bar b \bar c}{}_{\bar a} h^a{}_{\bar b} h_{a \bar c} \nonumber \\ && - 4 d \left( \bar \nabla^a \bar \nabla^{\bar a} h_{a \bar a} + \bar \nabla^{\bar a} \bar \nabla^a h_{a \bar a}\right) - 4 d \left( \bar \nabla^a \bar \nabla_a d  - \bar \nabla^{\bar a} \bar \nabla_{\bar a} d \right) \ ,
\end{eqnarray}
\begin{eqnarray}
{\cal L}_3 \!\!\!&=&\!\!\! - 4\, h_{a \bar b} \left(\bar \nabla^a h^{b \bar a} \bar \nabla^{\bar b}h_{b \bar a} - \bar \nabla^a h_{b \bar a} \bar \nabla^{\bar a} h^{b \bar b} - \bar \nabla^{\bar b} h_{b \bar a} \bar \nabla^b h^{a \bar a}  + \frac 1 2 \left(\bar \nabla^{[a}\bar \nabla^{\bar a]}h_{b \bar a} h^{b \bar b} - \bar \nabla^{[b}\bar \nabla^{\bar b]}h_{b \bar a} h^{a \bar a}\right)\right) \nonumber\\ &&\!\!\! - 2\, {\bar {\cal R}}^{a b \bar a c} {h}_{b \bar a} {h}_{a}\,^{\bar b} {h}_{c \bar b} - 2\, {\bar {\cal R}}^{\bar a \bar b a \bar c} {h}_{a \bar b} {h}^{b}\,_{\bar a} {h}_{b \bar c} \nonumber \\ &&\!\!\!
- 4 d \bigg[ \bar \nabla^a h_{a \bar a} \bar \nabla_b h^{b \bar a} - \bar \nabla^{\bar a} h_{a \bar a} \bar \nabla_{\bar b} h^{a \bar b} + \frac 1 2 \left( \bar \nabla^a h^{b \bar a} \bar \nabla_{a} h_{b \bar a} - \bar \nabla^{\bar a} h^{a \bar b} \bar \nabla_{\bar a} h_{a \bar b} \right)\nonumber \\
 && \!\!\!\ \ \ \ \ \ \  +\, 2 h_{a \bar b} \left( \bar \nabla^{(a} \bar \nabla^{b)} h_b{}^{\bar b} - \bar \nabla^{(\bar a}\bar \nabla^{\bar b)} h^a{}_{\bar a}\right) +  {\bar {\cal R}}^{a b c}\,_{a}  {h}_{b}\,^{\bar a} {h}_{c \bar a}-   {\bar {\cal R}}^{\bar a \bar b \bar c}\,_{\bar a}  {h}^{a}\,_{\bar b} {h}_{a \bar c}\bigg] \nonumber \\ &&\!\!\!+ 8\, d \,  h_{a \bar a} \left( \bar \nabla^a \bar \nabla^{\bar a} d + \bar \nabla^{\bar a} \bar \nabla^a d \right)
 +\, 4\, d^2\, (\bar \nabla^a \bar \nabla_a d - \bar \nabla^{\bar a} \bar \nabla_{\bar a} d) \ .
\end{eqnarray}

\section{WZW $S^3$ background with $H$-flux}

For the sake of concreteness, in this appendix we give the details for the simplest WZW model based on $G = SU(2)$, corresponding to a background given by  $S^3$ with $H$-flux. First, we introduce the Pauli matrices
\be
 \sigma_1 = \left(\begin{matrix}0&1\\1&0 \end{matrix}\right) \ , \ \ \ \sigma_2 = \left(\begin{matrix} 0&-i\\i&0\end{matrix}\right) \ , \ \ \ \sigma_3 = \left(\begin{matrix}1&0\\0&-1 \end{matrix}\right) \ , \ \ \
\ee
which define the Lie algebra generators as
\be
t_a \ = \ t_{\bar a} \ = \ \tfrac 1 2 (\sigma_1 , \sigma_2 , \sigma_3) \ .
\ee
Moreover, given two elements $X,Y \in {\frak g}$, the explicit representation for the quadratic Cartan-Killing
form reads
\be
\langle X , Y \rangle = - 4 \, {\rm Tr} ( X \, Y ) \ .
\ee
We can now compute the structure constants (with lowered indices)
\begin{eqnarray}
f_{a b c} &=& \langle [t_{a} , t_{b}] , t_{c}\rangle \ = \ - 2 i\,  \epsilon_{abc} \;, \\
f_{\bar a \bar b \bar c} &=& \langle [t_{\bar a} , t_{\bar b}] , t_{\bar c}\rangle \ = \ - 2 i \, \epsilon_{\bar a \bar b \bar c}  \ ,
\end{eqnarray}
where $\epsilon_{123} = \epsilon_{\bar 1 \bar 2 \bar 3} = 1$. We recall that at this stage there
is no difference between unbarred and barred indices, and so the second equation above strictly speaking is redundant.
Similarly, the Killing metric reads
\begin{eqnarray}
\kappa_{a b} &=& \langle t_a , t_b \rangle \ = \ - 2\, {\rm diag}(1,1,1) \;,
\end{eqnarray}
with the identical form for $\kappa_{\bar{a}\bar{b}}$.
Note that $\kappa_{a b} = -f_{ac}{}^{d} f_{bd}{}^{c}$.

We next introduce coordinates on the group manifold of $SU(2)$. To this end we use that
a general group element $\gamma$ can be parametrized as
\be\label{gammaMat}
\gamma \ = \ y_0{\bf 1} + i y_1 \sigma_1 + i y_2 \sigma_2 + i y_3 \sigma_3 \ ,
\ee
provided
\be
y_0^2 + y_1^2 + y_2^2 + y_3^2 \ = \ 1 \ .
\ee
Indeed, one may quickly verify with
$\{\sigma_i,\sigma_j\}=2\delta_{ij}$ that then $\gamma\gamma^{\dagger}={\bf 1}$.
We can view $(y_0,y_i)$ as coordinates for
$\mathbb{R}^4$, which are constrained by this relation to describe $S^3$.
For completeness we mention that
these constrained coordinates are related to the so-called Hopf coordinates $(\eta_1 , \eta_2 , \eta_3)$, where $\eta_1 \in (0 ,\, \frac \pi 2)$ and $\eta_2 ,\eta_3 \in (0 ,\, 2\pi)$,  by the relations
\begin{eqnarray}
y_0 &=& \cos \eta_2 \cos\eta_1 \ , \ \ \ \ y_1 \ = \ \sin \eta_2 \cos\eta_1\;,  \\
y_2 &=& \cos \eta_3 \sin\eta_1 \ , \ \ \ \ y_3 \ = \ \sin \eta_3 \sin \eta_1  \ .\nonumber
\end{eqnarray}

We can now determine the left- and right-invariant Maurer-Cartan forms (i.e.~the vielbeine), assuming that
the group element is given by (\ref{gammaMat}). A direct computation gives
\begin{eqnarray}
e_i{}^{a} &=& \langle \partial_i \gamma \gamma^{-1} , t^a \rangle  \ = \ 2 i \left(\begin{matrix} 0 &  \cos\eta^-_{23} & - \sin \eta^-_{23} \\  \cos^2 \eta_1 &  \cos \eta_1 \sin \eta_1 \sin \eta^-_{23} & \cos \eta_1 \sin \eta_1 \cos \eta^-_{23} \\ - \sin^2 \eta_1 &  \cos \eta_1 \sin \eta_1 \sin \eta^-_{23} &  \cos \eta_1 \sin \eta_1 \cos \eta^-_{23}\end{matrix}\right)\,, \label{eiaR} \\
\bar e_i{}^{\bar a} &=& \langle \gamma^{-1} \partial_i \gamma , t^{\bar a} \rangle \ = \ 2i  \left(\begin{matrix} 0 &  \cos\eta^+_{23} &  \sin \eta^+_{23} \\  \cos^2 \eta_1 &  \cos \eta_1 \sin \eta_1 \sin \eta^+_{23} & - \cos \eta_1 \sin \eta_1 \cos \eta^+_{23} \\  \sin^2 \eta_1 & - \cos \eta_1 \sin \eta_1 \sin \eta^+_{23} &  \cos \eta_1 \sin \eta_1 \cos \eta^+_{23}\end{matrix}\right)  \; \label{eiaL}
\end{eqnarray}
where $\eta^\pm_{23} = \eta_2 \pm \eta_3$. The representative of $\gamma$ in the adjoint representation
is according to (\ref{adjoint}) given by
\be
\gamma_{a}{}^{\bar a} \ = \ \langle t_a\, ,\, \gamma\, t^{\bar a}\, \gamma^{-1} \rangle \ .
\ee
It is then straightforward to verify explicitly that $\bar e_i{}^{\bar a} = e_i{}^b  \gamma_b{}^{\bar a}$ and that $\gamma$
is an element of $SO(3)$, i.e.,  $\kappa_{a b} = \gamma_a{}^{\bar a} \kappa_{\bar a \bar b} \gamma_b{}^{\bar b}$ and $\det \gamma = 1$.

Let us next turn to the two-form $B_{ij}$, which should be defined so that its field strength is
given by (c.f.~(\ref{HFORM}))
\be
H_{ijk} \ = \ - e_i{}^a e_j{}^b e_k{}^c f_{abc} \ \ \to \ \ H_{123} \ = \ - 16 \cos \eta_1 \sin \eta_1 \ .
\ee
This determines $B_{ij}$ up to gauge transformations. A convenient choice is
\be
B_{23} \ = \ 8 \cos^2 \eta_1
\ .
\ee

We have now all the data in order to define the background generalized frame
 \be
  \bar{E}_{A}{}^{M} \ = \  \begin{pmatrix}  e_{ia} +B_{ij}e_{a}{}^{j} & e_{a}{}^{i} \\[0.5ex]
  -\bar{e}_{i\bar{a}}+B_{ij}\bar{e}_{\bar{a}}{}^{j} & \bar{e}_{\bar{a}}{}^{i} \end{pmatrix}\ .
 \ee
A direct computation yields the background flat metric
\be
\bar {\cal G}_{A B} = \bar E_A{}^M \bar E_{B M} = \left(\begin{matrix} 2 \kappa_{a b} & 0 \\ 0 & - 2 \kappa_{\bar a \bar b}\end{matrix}\right) \ , \label{FlatMetS3Hflux}
\ee
in agreement with (\ref{BackgroundG}). Moreover, the fluxes (\ref{FDef}) are computed to be
\be
F_{ABC} \ = \ \left(\widehat{\cal L}_{\bar E_A} \bar E_B{}^M\right) \bar E_{C M} \ = \ \left\{\begin{matrix}F_{abc} = 4 i\,  \epsilon_{abc}    \\ F_{a b \bar c} = 0 \ \ \ \ \ \  \\ F_{a \bar b \bar c} = 0 \ \ \ \ \ \    \\ F_{\bar a \bar b \bar c} = 4 i\,  \epsilon_{\bar a \bar b \bar c} \end{matrix} \right. \ . \label{FluxesS3Hflux}
\ee
This confirms the result of the general analysis (\ref{structurconst}),
\be
F_{abc} \ = \ -2 f_{abc} \ , \ \ \ \ F_{\bar a \bar b \bar c} \ = \ - 2 f_{\bar a \bar b\bar c} \ .
\ee
In addition, taking into account the vanishing background value for the dilaton, $\phi = 0$, one can also compute the
trace part of the connections/fluxes,
\be
F_A = \partial_M \bar E_A{}^M - 2 D_A \bar d = 0\;,
\ee
in agreement with (\ref{FAiszero}).
Finally, one may also verify that the background DFT equations are satisfied,
as it should be in view of the general  results in sec. 4.

We close this appendix with a brief discussion of the conventional presentation of WZW backgrounds
and in particular with the standard conventions for the $S^3$ with $H$-flux. First, in order to treat a sphere
of general radius, we introduce
a dimensionful parameter $\rho$ and perform a rescaling of the frame and the two-form,
\be
e_i{}^{a} \;\to\; \frac {\rho} {\sqrt{8}} e_i{}^{a} \ , \ \ \ \ \bar e_i^{\bar a} \;\to\; \frac {\rho} {\sqrt{8}} \bar e_i{}^{\bar a} \ , \ \ \ \ B_{i j} \;\to\; \frac {\rho^2} {8} B_{i j} \ .
\ee
From the point of view of the WZW model the new parameter $\rho$
is related to $\alpha'$ and the level $k$ of the Kac-Moody algebra through $\rho = \sqrt{\alpha' k}$. The field theory limit corresponds to the large $\rho$ limit, which in turn implies large $k$.
After the rescaling the solution takes the form
\be
g_{ij} = \rho^2\, {\rm diag}(1 , \cos^2\eta_1 , \sin^2 \eta_1) \ , \ \ \ B_{23} = \rho^2 \cos^2 \eta_1 \ , \ \ \ \phi = 0 \ . \label{S3Hflux}
\ee
The Ricci tensor is given by $R_{ij} = 2\, {\rm diag} (1 , \cos^2\eta_1 , \sin^2 \eta_1)$ so the Ricci scalar gives $R = \frac 6 {\rho^2}$. In addition, one finds that $H^2 = \frac {24} {\rho^2}$, and therefore
\be
R + 4 (\partial \phi)^2 - \frac 1 {12} H^2 = \frac 4 {\rho^2} \ . \label{anomalousCC}
\ee
This implies that (\ref{S3Hflux}) is not a solution to the equations of motion of supergravity. In supergravity one way to solve this problem is to give the dilaton a linear dependence on time (which dimension we so far suppressed), $\phi = - \frac t {\rho}$ \cite{Callan:1991at}. Then, taking the time component of the metric as $g_{tt} = - 1$ one gets $4 (\partial \phi)^2 = - \frac 4 {\rho^2}$ which cancels the anomalous contribution (\ref{anomalousCC}). Here we intend to avoid this dilaton behavior, as it is not required by the WZW worldsheet CFT.

From the DFT perspective, after the rescaling, the background EOMs take the form
\be
\bar {\cal R} + \lambda = \frac 4 {\rho^2} + \lambda = 0 \ , \ \ \ \ \ \bar {\cal R}_{a \bar b} = 0 \,,
\ee
which is easily computed recalling that $\bar {\cal G}_{AB}$ remains unchanged after the rescaling (\ref{FlatMetS3Hflux}), that $F_A = 0$, and that the fluxes are now given by $F_{a b c} = i \frac {8 \sqrt{2}} {\rho} \epsilon_{abc}$ and  $F_{\bar a \bar b \bar c} = i \frac {8 \sqrt{2}} {\rho} \epsilon_{\bar a \bar b \bar c}$.
We then find complete agreement for a vanishing cosmological constant with the above computation
in standard geometry. Thus, in order to obtain a consistent CFT background we have to
include a cosmological constant, corresponding to string theory in a non-critical dimension. Indeed, following \cite{Nepomechie:1985ze}, one can identify $\lambda = -\frac 2 {3 \alpha'} (D-26)$,
\be
\bar {\cal R} + \lambda = \frac 4 {\rho^2} - \frac {2 (D - 26)} {3 \alpha'} = 0 \ .
\ee
Here $D = d + n$ encodes the sum of $d$ flat Minkowski directions plus the dimension of the group, $n = {\rm dim}(G)$.
Here, the gauge group is $G = SU(2)$, so $n = 3$. Generally, coupling flat space  to a WZW model with gauge group $SU(N)$, the flat critical dimension is given by \cite{Nemeschansky:1984mr}
\be
d - 26 = - \frac {(N^2 - 1)k}{N + k} = - (N^2 - 1) + \frac {N (N^2 - 1)}{k} + {\cal O}(k^{-2}) \ ,
\ee
where we performed a $\frac{1}{k}$ expansion.
When $N = 2$, this becomes to leading order
\be
d - 26 = - 3 + \frac 6 k \ ,
\ee
which is in agreement with $\lim_{k \to \infty} d = 23$. Then, on the background
\be
\bar {\cal R} + \lambda = \frac 4 {k \alpha'} - \frac {2 (3 + d - 26)}{3 \alpha'} = \frac 4 {k \alpha'} - \frac {2 (3  - 3 + \frac 6 k)}{3 \alpha'} = 0 \ ,
\ee
and the apparent anomaly cancels exactly.

\end{appendix}

\end{document}